\documentclass[12pt,preprint]{aastex63}
\usepackage{natbib}
\usepackage{amsmath}
\input{epsf}
\shorttitle{NIRI Imaging of Massive Stars in Cyg OB2}
\shortauthors{Caballero-Nieves et al.}
\begin{document}


\title{A High Angular Resolution Survey of Massive Stars in Cygnus~OB2: 
$JHK$ Adaptive Optics Results from the \\ Gemini Near-InfraRed Imager}

\author{S.~M. Caballero-Nieves}
\affiliation{Department of Physics and Space Sciences, Florida
  Institute of Technology 150 University Blvd, Melbourne, FL 32901, USA},
\author{D.~R. Gies}
\affiliation{Center for High Angular Resolution Astronomy,
Department of Physics and Astronomy, 
Georgia State University, P. O. Box 5060, Atlanta, GA  30302-5060, USA},
\author{E.~K. Baines}
\affiliation{Remote Sensing Division, Naval Research Laboratory,
4555 Overlook Avenue SW, Washington, DC 20375, USA},
\author{A.~H. Bouchez}
\affiliation{GMTO Corporation, 251 South Lake Avenue, Pasadena, CA
  91101,USA},
\affiliation{Observatories of the Carnegie Institution for Science,
813 Santa Barbara Street, Pasadena, CA 91101, USA},
\author{R.~G. Dekany}
\affiliation{California Institute of Technology, 1200 East
California Boulevard, Pasadena, CA 91125, USA},
\author{S.~P. Goodwin}
\affiliation{Department of Physics and Astronomy, University of
  Sheffield, Hicks Building, Hounsfield Road, Sheffield, S3 7RH,
  United Kingdom},
\author{E.~L. Rickman}
\affiliation{Geneva Observatory, University of Geneva, Chemin des Maillettes
  51, CH-1290 Sauverny, Switzerland},
\author{L.~C. Roberts Jr.}
\affiliation{Jet Propulsion Laboratory, California Institute of
Technology, 4800 Oak Grove Drive, Pasadena, CA 91109, USA},
\author{K.~Taggart}
\affiliation{Astrophysics Research Institute, Liverpool John
  Moores University, IC2, Liverpool Science Park, 146 Brownlow Hill,
  Liverpool, L3 5RF, UK},
\author{T.~A. ten Brummelaar}
\affiliation{The CHARA Array, Mount Wilson Observatory, Mount
Wilson, CA 91023, USA},
\author{N.~H. Turner}
\affiliation{The CHARA Array, Mount Wilson Observatory, Mount
Wilson, CA 91023, USA}


\journalinfo{\today}


\begin{abstract}
We present results of a high angular resolution survey of massive OB
stars in the Cygnus OB2 association that we conducted with the NIRI
camera and ALTAIR adaptive optics system of the Gemini North
telescope.  We observed 74 O- and early B-type stars in Cyg OB2 in the $JHK$
infrared bands in order to detect binary and multiple companions.
The observations are sensitive to equal-brightness pairs at 
separations as small as $0\farcs08$, and progressively fainter 
companions are detectable out to $\triangle K = 9$ mag at 
a separation of $2 \arcsec$.  This faint contrast limit due to 
readnoise continues out to 10 arcsec near the edge of the detector.
We assigned a simple probability of chance alignment to each 
companion based upon its separation and magnitude difference from
the central target star and upon areal star counts for the general
star field of Cyg~OB2. Companion stars with a field membership 
probability of less than $1\%$ are assumed to be physical companions. 
This assessment indicates that $47\%$ of the targets have at least one
resolved companion that is probably gravitationally bound.
Including known spectroscopic binaries, our sample includes 27 binary,
12 triple, and 9 systems with four or more components.  These results
confirm studies of high mass stars in other environments that find that massive
stars are born with a high multiplicity fraction.  The results are
important for the placement of the stars in the H-R diagram, the
interpretation of their spectroscopic analyses, and for future mass
determinations through measurement of orbital motion.
\end{abstract}

\keywords{techniques: high angular resolution 
--- binaries: visual
--- stars: early-type 
--- stars: massive 
--- open clusters and associations: individual: Cyg OB2}


\section{Introduction}                              

Massive stars profoundly influence the evolution of the Universe, 
from galactic dynamics and structure to star formation. 
They are often found with bound companions. 
Massive stars have a higher frequency of multiplicity than cooler, 
less massive stars \citep{rag10,duc13}, 
especially when they are found in clusters \citep{mas09}. 
Spectroscopic studies of massive stars in the Milky Way  
\citep{san12} and in the Tarantula Nebula region of the Large 
Magellanic Cloud \citep{san13} demonstrate that perhaps $75\%$ of 
massive O-type stars have binary companions in orbits small 
enough that the stars will interact over their lifetime. 
Our knowledge of O-type multiple systems in larger orbits 
with periods in the range from years to thousands of years is 
incomplete due their great distances, but high angular resolution
methods are beginning to fill in this period gap \citep{mai19,leb17,ald15,san14}.  

At a distance of 1.33 -- 1.7~kpc \citep{mas91,tor91,han03,ryg12,kim15},
Cygnus~OB2 = Cyg~OB2 is the second closest OB association (after Ori OB1) 
that provides us with an example of a nearby, young stellar environment, 
rich in high-mass stars.  Due to uneven extinction towards the region 
\citep{wri15}, the cluster begins to be unveiled in the infrared (IR). 
\citet{tor91} estimate the age of the association to be least
3 Myr through analysis of their Str\"{o}mgren and infrared photometry,
and \citet{wri15} argue that star formation has occurred more or less 
continuously over the last 1 to 7 Myr based upon the positions of the
stars in the Hertzsprung - Russell (H-R) diagram.  The young nature of
the association is further established by the detection of X-rays 
from young, low mass stars in the region \citep{alb07,wri09,wri12}. 
Spectroscopic surveys by \citet{mas91}, \citet{han03}, and \citet{kim07} 
have established the early-type classifications of these stars. 
Massive stars are short-lived and therefore spend most of 
their formative years shrouded in their natal clouds, so that when 
they shed these clouds and a hot star is revealed, it is usually
well into the main sequence stage of its life.  The multiplicity of 
massive stars must play an important role in their formation because 
so many are members of binary systems \citep{zin07}. Massive stars are formed 
through the turbulent core collapse of a single cloud or by 
competitive accretion of multiple stellar seeds in a dense cloud 
(see the review by \citealt{ros20}), and models of these processes 
predict a large incidence of binary stars with specific distributions 
of mass ratio and separation \citep{pet12,gra18}. Therefore,
by studying the multiplicity properties of massive stars at an
early stage we can test the role of companions in formation theories 
of massive stars.  

The Cyg OB2 association is close enough that with modern-day adaptive
optics (AO) we are able to resolve relatively close companions.  The
ALTAIR AO system and the Near-InfraRed Imager and Spectrograph (NIRI)
at the Gemini North Observatory provides an effective tool to search
for binaries, as was demonstrated by \citet{laf14} in a multiplicity
study of young stars in the Upper Sco association.  With a resolution
of $\sim 0\farcs06$ and a sensitivity contrast limit of about 10~mag
for differential photometry, the ALTAIR AO infrared system can delve
into the depths of the association and find faint companions with
periods in the range from hundreds to thousands of years.  Our results
complement the radial velocity survey of \citet{kob14} (and references
therein) who searched for short period, spectroscopic systems in
Cyg~OB2. They determined that $30\%$ of their sample are spectroscopic
binaries with periods less than 45 days.

In this paper, we provide measurements of $JHK$-band relative photometry
and positions of candidate companions to our target stars. 
These results provide the first step in determining the true 
multiplicity fraction of widely separated systems. 
In section 2 we describe the observations of the sample in Cygnus OB2. 
We present the results of the survey in section 3 along with further 
details of the calibration in appendices
for the astrometry (Appendix A) and photometry (Appendix B). 
We discuss the detection limits and the identification
of probable physical companions in section 4.  Section 5 presents 
the multiplicity fraction and companion frequency for the Cyg~OB2 
sample and compares these to similar results from studies of 
massive stars in other locations.  We summarize the results and
their significance in section 6.


\section{Observations}\label{niridata}            

We were able to observe 74 of the brightest O- and B-type stars in
Cyg~OB2 and one misidentified non-member using the infrared ALTAIR AO
system \citep{ric98, sin98} at the Gemini North Observatory.  We
provide a list of our targets in Table \ref{t1-niri} (given in full in
the electronic version) that gives the target name, celestial
coordinates (J2000), spectral classification and reference, optical
and infrared (IR) magnitudes, and three measures of interstellar
reddening.  The majority of stars in this study were selected from the
optical survey of \citet{mas91}, who presented Johnson $B$ and $V$
magnitudes for the brighter stars in the sample as well as reddening
towards each star.  These targets are identified with prefix ``MT'' by
their number assigned by \citet{mas91}.  Seventeen of our targets were
selected from the infrared surveys by \citet{com02,com08}, and these
are referenced by a prefix ``A'' or ``B'' from those papers.  These
are redder sources that are not readily detected in the optical
surveys, but $V$-band and spectral information are available for some
of these from \cite{str08}.  An ``S'' designation is given for three
stars from the compilation by \citet{sch58}, and the final object is
given by its Wolf-Rayet catalog number \citep{van01}.  The spectral
classifications are taken from a variety of sources, indicated in the
notes below the table. The $V$ magnitudes reported are from
\citet{mas91} for the MT~\# stars and from \citet{str08} for
others. The coordinates and infrared $JHK_s$ photometry are from the
Two Micron All Sky Survey (2MASS; \citealt{skr06}).  The reddening
estimates are discussed in section 4.3 below. After we completed our
observations we learned that the object MT~140 is in fact an
intermediate mass object that is not a member of Cyg~OB2
\citep{mai16}. We include our measurements here for completeness, but
it is excluded from the multiplicity analysis.

\citet{wri15} describe the massive star content of the 
Cyg~OB2 association, and they suggest that the association 
hosts 52 O-type stars and 3 Wolf-Rayet stars. Our sample 
includes 56 O-type stars and one Wolf-Rayet star, plus 
a number of luminous and/or early type B-stars.  Thus, 
our target list should represent an almost complete sample 
of the most massive stars ($M/M_\odot > 18$) in Cyg~OB2
(missing the Wolf-Rayet Stars WR~144 and WR~146). 


\begin{longrotatetable}
\begin{deluxetable}{lcclccccccccc}
\tablewidth{0pc}
\tabletypesize{\scriptsize}
\setlength{\tabcolsep}{0.1in} 
\rotate
\tablenum{1}
\tablecaption{Sample of Stars in Cygnus OB2\label{t1-niri}}
\tablewidth{0pt}
\tablehead{
\colhead{Star} &
\colhead{Schulte \#} &
\colhead{R.A.} &
\colhead{Dec.} &
\colhead{Spectral} &
\colhead{Class.} &
\colhead{$V$} &
\colhead{$J$} &
\colhead{$H$} &
\colhead{$K_s$} &
\colhead{$E(J-K)$} & 
\colhead{$E(B-V)$} &
\colhead{$E(b-y)$} \\
\colhead{Name} &
\colhead{} &
\colhead{(HH:MM:SS)} &
\colhead{(DD:MM:SS)} &
\colhead{Classification} &
\colhead{Ref.} &
\colhead{(mag)} &
\colhead{(mag)} &
\colhead{(mag)} &
\colhead{(mag)} &
\colhead{(mag)} &
\colhead{(mag)} &
\colhead{(mag)} }
\startdata
A~11       & (MT~267)   & 20:32:31.539 & +41:14:08.22 & O7.5 III-1    & 1 & \nodata &  7.817 &  7.094 &  6.664 & 1.32    & \nodata & \nodata \\
A~12       &           & 20:33:38.219 & +40:41:06.41 & B0 Ia         & 3 & \nodata &  6.904 &  6.170 &  5.745 & 1.40    & \nodata & \nodata \\
A~15       &           & 20:31:36.909 & +40:59:09.25 & O7 Ib(f)      & 3 & \nodata &  7.913 &  7.208 &  6.811 & 1.32    & \nodata & \nodata \\
A~18       &           & 20:30:07.879 & +41:23:50.47 & O8 V          & 3 & \nodata &  9.397 &  8.739 &  8.365 & 1.25    & \nodata & \nodata \\
A~20       &           & 20:33:02.920 & +40:47:25.45 & O8 II((f))    & 5 & \nodata &  7.251 &  6.632 &  6.274 & 1.16    & \nodata & \nodata \\
A~23       &           & 20:30:39.710 & +41:08:48.98 & B0.7          & 5 & \nodata &  6.928 &  6.328 &  5.980 & 1.08    & \nodata & \nodata \\
A~24       &           & 20:34:44.110 & +40:51:58.51 & O6.5 III((f)) & 3 & \nodata &  8.405 &  7.796 &  7.448 & 1.15    & \nodata & \nodata \\
A~25       &           & 20:32:38.441 & +40:40:44.54 & O8 III        & 3 & \nodata &  8.347 &  7.705 &  7.383 & 1.19    & \nodata & \nodata \\
A~26       &           & 20:30:57.730 & +41:09:57.57 & O9.5 V        & 3 & \nodata &  9.093 &  8.514 &  8.198 & 1.14    & \nodata & \nodata \\
A~27       &           & 20:34:44.719 & +40:51:46.56 & B0 Ia         & 3 & \nodata &  6.683 &  6.062 &  5.731 & 1.14    & \nodata & \nodata \\
A~29       &           & 20:34:56.061 & +40:38:18.06 & O9.7 Iab      & 5 & \nodata &  7.440 &  6.859 &  6.545 & 1.05    & \nodata & \nodata \\
A~32       &           & 20:32:30.330 & +40:34:33.26 & O9.5 IV       & 5 & \nodata &  7.892 &  7.365 &  7.070 & 1.01    & \nodata & \nodata \\
A~37       &           & 20:36:04.520 & +40:56:12.98 & O5 V          & 5 & \nodata &  8.568 &  7.968 &  7.685 & 1.04    & \nodata & \nodata \\
A~38       &           & 20:32:34.870 & +40:56:17.42 & O8 V          & 3 & \nodata &  9.382 &  8.858 &  8.564 & 0.98    & \nodata & \nodata \\
A~41       &           & 20:31:08.378 & +42:02:42.28 & O9.7 II       & 5 & \nodata &  7.828 &  7.292 &  7.023 & 0.96    & \nodata & \nodata \\
A~46       &           & 20:31:00.200 & +40:49:49.75 & O7 V          & 5 & \nodata &  8.378 &  8.016 &  7.826 & 0.70    & \nodata & \nodata \\
B~17       &           & 20:30:27.299 & +41:13:25.31 & O7:           & 1 & \nodata &  7.630 &  6.850 &  6.445 & \nodata & \nodata & \nodata \\
MT~5       &           & 20:30:39.820 & +41:36:50.72 & O6 V          & 2 &   12.93 &  9.098 &  8.574 &  8.313 & 0.95    & 1.96    & \nodata \\
MT~59      & CygOB2-1  & 20:31:10.549 & +41:31:53.55 & O8 V          & 1 &   11.06 &  7.968 &  7.556 &  7.365 & 0.76    & 1.78    & 1.20    \\
MT~70      &           & 20:31:18.330 & +41:21:21.66 & O9 II         & 1 &   12.99 &  8.607 &  8.046 &  7.746 & 1.04    & 2.41    & \nodata \\
MT~83      & CygOB2-2  & 20:31:22.038 & +41:31:28.41 & B1 I          & 2 &   10.61 &  8.075 &  7.750 &  7.628 & 0.58    & 1.37    & 1.01    \\
MT~138     &           & 20:31:45.400 & +41:18:26.75 & O8 I          & 2 &   12.26 &  8.065 &  7.552 &  7.259 & 0.99    & 2.27    & 1.49    \\
MT~140\tablenotemark{*}&           & 20:31:46.011 & +41:17:27.07 & F             & 5 &    9.38 &  8.240 &  8.061 &  8.048 & \nodata & \nodata & \nodata \\
MT~145     & CygOB2-20 & 20:31:49.659 & +41:28:26.52 & O9 III        & 1 &   11.62 &  9.074 &  8.768 &  8.634 & 0.62    & 1.41    & 0.99    \\
MT~213     &           & 20:32:13.130 & +41:27:24.63 & B0 V          & 2 &   11.95 &  9.521 &  9.248 &  9.071 & 0.63    & 1.43    & \nodata \\
MT~217     & CygOB2-4  & 20:32:13.830 & +41:27:12.03 & O7 IIIf       & 2 &   10.07 &  7.582 &  7.248 &  7.105 & 0.67    & 1.50    & 1.03    \\
MT~227     & CygOB2-14 & 20:32:16.560 & +41:25:35.71 & O9 V          & 2 &   11.47 &  8.714 &  8.389 &  8.185 & 0.71    & 1.55    & 1.06    \\
MT~250     &           & 20:32:26.079 & +41:29:39.36 & B2 III        & 2 &   12.88 & 10.427 & 10.150 &  9.993 & 0.61    & 1.32    & \nodata \\
MT~258     & CygOB2-15 & 20:32:27.660 & +41:26:22.08 & O8 V          & 1 &   10.90 &  8.535 &  8.193 &  8.021 & 0.67    & 1.51    & 1.04    \\
MT~259     & CygOB2-21 & 20:32:27.739 & +41:28:52.28 & B0 Ib         & 2 &   11.50 &  9.191 &  8.895 &  8.766 & 0.57    & 1.28    & 0.90    \\
MT~299     & CygOB2-16 & 20:32:38.579 & +41:25:13.75 & O7 V          & 2 &   11.12 &  8.194 &  7.918 &  7.716 & 0.63    & 1.50    & 1.03    \\
MT~304     & CygOB2-12 & 20:32:40.958 & +41:14:29.16 & B3 Iae        & 2 &   11.40 &  4.667 &  3.512 &  2.704 & \nodata & 3.44  & \nodata   \\
MT~317     & CygOB2-6  & 20:32:45.458 & +41:25:37.43 & O8 V          & 2 &   10.65 &  7.953 &  7.617 &  7.421 & 0.69    & 1.56    & 1.07    \\
MT~339     & CygOB2-17 & 20:32:50.019 & +41:23:44.68 & O8 V          & 2 &   11.71 &  8.579 &  8.188 &  7.982 & 0.76    & 1.66    & 1.15    \\
MT~376     &           & 20:32:59.190 & +41:24:25.50 & O8 V          & 2 &   11.91 &  8.886 &  8.524 &  8.314 & 0.73    & 1.66    & 1.13    \\
MT~390     &           & 20:33:02.920 & +41:17:43.14 & O8 V          & 2 &   12.95 &  8.718 &  8.165 &  7.873 & 1.01    & 2.29    & 1.51    \\
MT~403     &           & 20:33:05.269 & +41:43:36.80 & B1 V          & 2 &   12.94 &  9.286 &  8.854 &  8.624 & 0.81    & 1.74    & \nodata \\
MT~417     & CygOB2-22 & 20:33:08.801 & +41:13:18.21 & O3 I          & 6 &   11.68 &  7.110 &  6.540 &  6.226 & 1.08    & 2.36    & 1.60    \\
MT~421     & CygOB2-50 & 20:33:09.600 & +41:13:00.54 & O9 V          & 1 &   12.86 &  8.655 &  8.135 &  7.764 & \nodata & 2.26    & 1.50    \\
MT~429     &           & 20:33:10.508 & +41:22:22.44 & B0 V          & 1 &   12.98 &  9.537 &  9.113 &  8.897 & 0.82    & 1.86    & \nodata \\
MT~431     & CygOB2-9  & 20:33:10.751 & +41:15:08.20 & O5:           & 1 &   10.78 &  6.468 &  5.897 &  5.570 & 1.09    & 2.11    & 1.52    \\
MT~448     &           & 20:33:13.258 & +41:13:28.74 & O6 V          & 2 &   13.61 &  8.982 &  8.346 &  8.009 & 1.13    & 2.47    & \nodata \\
MT~455     &           & 20:33:13.690 & +41:13:05.79 & O8 V          & 2 &   12.92 &  9.034 &  8.559 &  8.280 & 0.91    & 2.12    & \nodata \\
MT~457     & CygOB2-7  & 20:33:14.110 & +41:20:21.81 & O3 If         & 2 &   10.50 &  7.248 &  6.818 &  6.611 & 0.83    & 1.76    & 1.23    \\
MT~462     & CygOB2-8B & 20:33:14.759 & +41:18:41.63 & O7 III-II     & 2 &   10.70 &  7.209 &  6.762 &  6.570 & 0.83    & 1.75    & 1.13    \\
MT~465     & CygOB2-8A & 20:33:15.079 & +41:18:50.45 & O5.5 I        & 1 &    8.99 &  6.123 &  5.721 &  5.503 & 0.81    & 1.60    & 1.09    \\
MT~470     & CygOB2-23 & 20:33:15.708 & +41:20:17.20 & O9 V          & 2 &   12.61 &  9.333 &  8.935 &  8.725 & 0.79    & 1.76    & 1.22    \\
MT~473     & CygOB2-8D & 20:33:16.338 & +41:19:01.80 & O8.5 V        & 2 &   12.02 &  8.842 &  8.424 &  8.239 & 0.78    & 1.76    & 1.14    \\
MT~480     & CygOB2-24 & 20:33:17.479 & +41:17:09.31 & O7 V          & 2 &   11.86 &  8.354 &  7.889 &  7.649 & 0.85    & 1.90    & 1.31    \\
MT~483     & CygOB2-8C & 20:33:17.989 & +41:18:31.10 & O5 III        & 2 &   10.08 &  7.165 &  6.792 &  6.579 & 0.79    & 1.54    & 1.11    \\
MT~485     &           & 20:33:18.030 & +41:21:36.65 & O8 V          & 2 &   11.82 &  8.744 &  8.315 &  8.113 & 0.79    & 1.82    & 1.25    \\
MT~507     &           & 20:33:21.020 & +41:17:40.14 & O9 V          & 2 &   12.70 &  9.301 &  8.899 &  8.672 & 0.81    & 1.85    & 1.39    \\
MT~516     &           & 20:33:23.458 & +41:09:13.00 & O5.5 V        & 2 &   11.84 &  7.025 &  6.380 &  6.050 & 1.14    & 2.52    & 1.75    \\
MT~531     & CygOB2-25 & 20:33:25.558 & +41:33:27.00 & O8.5 V        & 2 &   11.58 &  8.168 &  7.748 &  7.523 & 0.83    & 1.88    & 1.27    \\
MT~534     &           & 20:33:26.748 & +41:10:59.51 & O8.5 V        & 2 &   13.00 &  8.971 &  8.434 &  8.165 & 0.99    & 2.18    & \nodata \\
MT~555     & CygOB2-74 & 20:33:30.310 & +41:35:57.89 & O8 V          & 2 &   12.51 &  8.385 &  7.839 &  7.568 & 0.98    & 2.21    & \nodata \\
MT~556     & CygOB2-18 & 20:33:30.790 & +41:15:22.66 & B1 I          & 2 &   11.01 &  6.493 &  5.891 &  5.542 & 1.08    & 1.96    & 1.55    \\
MT~588     & CygOB2-70 & 20:33:37.000 & +41:16:11.30 & B0 V          & 2 &   12.40 &  8.683 &  8.168 &  7.929 & 0.93    & 1.96    & 1.40    \\
MT~601     & CygOB2-19 & 20:33:39.110 & +41:19:25.86 & B0 Iab        & 2 &   11.06 &  7.230 &  6.745 &  6.482 & 0.89    & 1.77    & 1.32    \\
MT~605     &           & 20:33:39.798 & +41:22:52.37 & B1 V          & 1 &   11.78 &  8.876 &  8.543 &  8.279 & 0.75    & 1.47    & 1.08    \\
MT~611     &           & 20:33:40.869 & +41:30:18.98 & O7 V          & 2 &   12.77 &  9.263 &  8.866 &  8.614 & 0.80    & 1.88    & 1.27    \\
MT~632     & CygOB2-10 & 20:33:46.100 & +41:33:01.05 & O9 I          & 2 &    9.82 &  6.294 &  5.839 &  5.582 & 0.87    & 1.86    & 1.28    \\
MT~642     & CygOB2-26 & 20:33:47.839 & +41:20:41.54 & B1 III        & 2 &   11.87 &  7.986 &  7.487 &  7.209 & 0.97    & 1.79    & 1.32    \\
MT~692     &           & 20:33:59.250 & +41:05:38.09 & B0 V          & 2 &   13.61 &  9.988 &  9.567 &  9.301 & 0.87    & 1.99    & \nodata \\
MT~696     & CygOB2-27 & 20:33:59.529 & +41:17:35.48 & O9.5 V        & 1 &   12.25 &  8.534 &  8.140 &  7.889 & 0.82    & 1.95    & 1.32    \\
MT~716     &           & 20:34:04.861 & +41:05:12.92 & O9 V          & 2 &   13.50 &  9.561 &  9.095 &  8.836 & 0.91    & 2.14    & \nodata \\
MT~734     & CygOB2-11 & 20:34:08.502 & +41:36:59.26 & O5 I          & 1 &   10.08 &  6.650 &  6.226 &  5.990 & 0.85    & 1.79    & 1.19    \\
MT~736     & CygOB2-75 & 20:34:09.520 & +41:34:13.70 & O9 V          & 2 &   12.79 &  9.304 &  8.892 &  8.646 & 0.84    & 1.77    & \nodata \\
MT~745     & CygOB2-29 & 20:34:13.509 & +41:35:02.74 & O7 V          & 2 &   12.04 &  8.550 &  8.148 &  7.921 & 0.78    & 1.82    & 1.26    \\
MT~771     &           & 20:34:29.600 & +41:31:45.55 & O7 V          & 1 &   11.64 &  7.560 &  7.030 &  6.709 & 1.00    & 2.37    & \nodata \\
MT~793     & CygOB2-30 & 20:34:43.580 & +41:29:04.63 & B2 IIIe       & 2 &   12.36 &  8.614 &  8.116 &  7.701 & 1.09    & 1.79    & 1.28    \\
Schulte~3  & CygOB2-3  & 20:31:37.501 & +41:13:21.04 & O6 IV:        & 1 &   10.35 &  6.498 &  6.001 &  5.748 & \nodata & \nodata & 1.36    \\
Schulte~5  & CygOB2-5  & 20:32:22.431 & +41:18:19.10 & O7 Ianfp      & 1 &    9.21 &  5.187 &  4.745 &  4.339 & \nodata & \nodata & 1.34    \\
Schulte~73 & CygOB2-73 & 20:34:21.929 & +41:17:01.60 & O8 III        & 1 &   12.50 &  8.388 &  7.878 &  7.602 & \nodata & \nodata & 1.45    \\
WR~145     &           & 20:32:06.289 & +40:48:29.54 & WN7o/CE       & 4 &   12.30 &  7.373 &  6.714 &  6.239 & \nodata & 2.03    & \nodata \\
\enddata
\tablerefs{
1. \citet{kob12};
2. \citet{kim07};
3. \citet{neg08};
4. \citet{mun09};
5. http://simbad.u-strasbg.fr/simbad/;
6. \citet{mas01}.}

\tablenotetext{*}{MT~140 appears to be an erroneous F-type star \citep{mai16}. We include the observations in the tables but this object is not included in the final analysis of MF and CF.}
\end{deluxetable}
\end{longrotatetable}

\clearpage

Our observations were made in three queue observing programs at the
8.1-m Gemini North Observatory during the 2005B, 2008A and 2008B
observing semesters. Using the Near InfraRed Imager and Spectrograph
(NIRI) with the ALTAIR adaptive optics (AO) system (\citealt{hod03}; \citealt{ric98}; \citealt{rob98}), we
collected high resolution images ($0\farcs022$~pixel$^{-1}$ with the
$f/32$ camera) with a field of view (FOV) of approximately $22\arcsec \times 22\arcsec$. 
The only exception is for our $K$-band observations of MT~304 = Cyg~OB2 \#12. 
Due to its extreme IR brightness ($K = 2.7$) MT~304 was observed with the 
shortest exposure time possible, and therefore, a smaller FOV ($11\arcsec \times
11\arcsec$) was used so that the data could be read out without over-exposing
the images. The detector chip used the deep well setting for improved
dynamic range, and the 2008 data were obtained with the ALTAIR field lens which
improves the AO correction. The telescope sits on an altitude-azimuth
mount, so that when NIRI is held fixed, the sky appears to rotate
between frames.  For these observations, NIRI was held fixed and the
exposure times for each frame ranged between 0.02~s to 800~s in $K$
and between 0.1~s to 1869~s in $J$, depending of the brightness of the
target star in each band in order to reach about half of the 
full well depth of the detector and achieve uniform S/N ratio 
measurements of the target stars.

Table~\ref{t2-niri} provides the central wavelength and the pass band
for each filter. Every target was observed with the $K$ continuum
filter, Kcon(209), to detect possible companions. The numbering
corresponds to the central wavelength in hundreds of angstroms. We followed up on
43 stars with $J$-band observations to get additional color
information on those systems with obvious companions. The 2005 data
were obtained using the $J$ continuum filter, Jcon(112). The wider
Jcon(121) filter was used for the 2008 observations because the
companions appear fainter in the $J$-band than in the $K$-band.  The
seven targets observed during the 2005B semester were also imaged with
the $H$ continuum filter, Hcon(157), with the exception of MT~304
which was only observed in $J$ at the time.  These filters all have
narrow pass bands that were needed because the stars are so bright in
the infrared.


\begin{deluxetable}{llcc}
\tablewidth{0pc}
\tablenum{2}
\tablecaption{Filter Information\label{t2-niri}}
\tablewidth{0pt}
\tablehead{
\colhead{Instrument} &
\colhead{Filter Name} &
\colhead{Central Wavelength} &
\colhead{Bandpass} \\
\colhead{} &
\colhead{} &
\colhead{($\mu$m)} &
\colhead{($\mu$m)} }
\startdata
NIRI  & Jcon(112) & 1.122  & 0.009 \\
NIRI  & Jcon(121) & 1.207  & 0.018 \\
NIRI  & Hcon(157) & 1.570  & 0.024 \\
NIRI  & Kcon(209) & 2.0975 & 0.027 \\
PHARO & J         & 1.246  & 0.162 \\
PHARO & H         & 1.635  & 0.296 \\
PHARO & K$_S$     & 2.145  & 0.310 \\
\enddata
\end{deluxetable}

Each observation consisted of approximately 90
frames. Table~\ref{t3-niri} (given in full in the electronic version)
lists the observation dates of the
beginning of the first exposure and the number of frames combined to
produce the final co-added image for each filter. Each target was
observed at nine dither positions, set up on a $3 \times 3$ grid,
offset by about 50 pixels and with 10 exposures at each
position.  For the cases where the observations were taken over
two nights, observations from each night were combined individually
and also combined together. For the detection of sources, the images
from each night were analyzed separately due to differences in image
quality, but only data from one night were used for photometric and
astrometric measurements (denoted by *). For A~25 in $K$ and A~41 in
$J$, we analyzed the combined image from both nights (denoted by $^C$)
because they were of comparable quality.  The fourth  and fifth columns
give the Strehl ratio and full-width at half-maximum (FWHM), respectively, 
of the point spread function associated with the primary target. 
These were determined using the IDL Strehl ratio meter 
code\footnote{http://www2.keck.hawaii.edu/optics/aochar/Strehl\_meter2.htm} 
written by M.\ van Dam. 

In addition to the NIRI $K$-band observation, MT~421 was observed
with the Palomar High Angular Resolution Optics (PHARO; \citealt{hay01}) camera and the Palm-3000 AO
system \citep{dek13} on the 5-m Hale telescope in 2009 July.  We were able to get observations in all three IR bands, $J$,
$H$, and $K_{S}$, with a field of view comparable to that of NIRI ($\sim
25\arcsec \times 25 \arcsec$). The filter information for PHARO is also
listed in Table~\ref{t2-niri}. The PHARO images provide a 
pixel scale of $0\farcs025$ pixel$^{-1}$ \citep{hay01}.


\startlongtable
\begin{deluxetable}{lccccc}
\tablewidth{0pc}
\tabletypesize{\scriptsize}
\tablenum{3}
\tablecaption{Observations of Stars in Cyg OB2\label{t3-niri}}
\tablewidth{0pt}
\tablehead{
\colhead{Star} &
\colhead{Date} &
\colhead{Filter} &
\colhead{Strehl} &
\colhead{FWHM} &
\colhead{Number of} \\
\colhead{Name} &
\colhead{(JD -- 2,450,000)} &
\colhead{Name} &
\colhead{Ratio} &
\colhead{(mas)} &
\colhead{Images} }
\startdata
A~11   & 4741.250 & Kcon(209)   & 0.32 & \phn 83 &  91 \\
A~12   & 4741.242 & Kcon(209)   & 0.32 & \phn 81 &  90 \\
A~15   & 4741.234 & Kcon(209)   & 0.33 & \phn 80 &  90 \\
A~18   & 4741.220 & Kcon(209)   & 0.28 & \phn 85 &  90 \\  
A~20   & 4741.210 & Kcon(209)   & 0.16 &     114 &  90 \\  
A~23   & 4590.621 & Kcon(209)   & 0.36 & \phn 76 &  90 \\  
A~24   & 4741.201 & Kcon(209)   & 0.33 & \phn 80 &  90 \\  
A~25   &4740.329$^{C}$&Kcon(209)& 0.15 &     124 &  69 \\  
       &4741.197$^{C}$&Kcon(209)& 0.15 &     124 &  22 \\  
A~26   & 4740.292 & Kcon(209)   & 0.19 &     102 &  90 \\  
A~27   & 4741.261 & Jcon(121)   & 0.05 &     105 &  89 \\  
       & 4593.612 & Kcon(209)   & 0.33 & \phn 77 &  90 \\  
A~29   & 4740.283 & Kcon(209)   & 0.31 & \phn 83 &  90 \\
A~32   & 4819.203 & Jcon(121)   & 0.01 &     159 &  90 \\  
       & 4740.273 & Kcon(209)   & 0.34 & \phn 81 &  90 \\  
A~37   & 4740.264 & Kcon(209)   & 0.37 & \phn 78 &  90 \\  
A~38   & 4740.249 & Kcon(209)   & 0.22 &     105 & 106 \\ 
A~41   &4742.207$^{C}$&Jcon(121)& 0.08 & \phn 82 &  60 \\  
       &4746.302$^{C}$&Jcon(121)& 0.08 & \phn 82 &  30 \\  
       & 4593.604 & Kcon(209)   & 0.34 & \phn 75 &  90 \\  
A~46   & 4593.594 & Kcon(209)   & 0.25 & \phn 90 &  90 \\  
B~17   & 4819.184 & Jcon(121)   & 0.02 &     142 &  90 \\  
       & 4740.241 & Kcon(209)   & 0.34 & \phn 81 &  90 \\ 
MT~5   & 4746.310*& Jcon(121)   & 0.03 &     115 &  90 \\
       & 4747.218 & Jcon(121)   & 0.03 &     115 &  90 \\  
       & 4603.605 & Kcon(209)   & 0.33 & \phn 81 &  90 \\  
MT~59  & 4743.249 & Jcon(121)   & 0.04 &     110 &  90 \\  
       & 4746.343*& Jcon(121)   & 0.04 &     110 &  90 \\  
       & 4607.571 & Kcon(209)   & 0.21 &     102 &  90 \\  
MT~70  & 4817.180 & Jcon(121)   & 0.04 &     114 &  90 \\  
       & 4607.580 & Kcon(209)   & 0.21 &     104 &  90 \\  
MT~83  & 4804.184 & Jcon(121)   & 0.04 &     105 &  90 \\  
       & 4598.615 & Kcon(209)   & 0.20 &     103 &  90 \\  
MT~138 & 4747.277 & Jcon(121)   & 0.06 &     102 &  90 \\  
       & 4607.590 & Kcon(209)   & 0.20 &     107 &  90 \\  
MT~140 & 4740.229 & Kcon(209)   & 0.22 &     107 &  63 \\
MT~145 & 4747.326 & Jcon(121)   & 0.06 &     101 & 130 \\ 
       & 4620.599 & Kcon(209)   & 0.33 & \phn 83 &  90 \\  
MT~213 & 4747.347 & Jcon(121)   & 0.06 & \phn 98 &  90 \\  
       & 4605.601 & Kcon(209)   & 0.23 &     101 &  90 \\  
MT~217 & 4818.176 & Jcon(121)   & 0.02 &     138 &  90 \\  
       & 4607.598 & Kcon(209)   & 0.17 &     114 &  90 \\  
MT~227 & 4607.607 & Kcon(209)   & 0.19 &     109 &  90 \\  
MT~250 & 4818.189 & Jcon(121)   & 0.04 & \phn 90 &  90 \\  
       & 4620.611 & Kcon(209)   & 0.35 & \phn 83 &  90 \\  
MT~258 & 4805.184 & Jcon(121)   & 0.03 &     127 &  90 \\  
       & 4607.617 & Kcon(209)   & 0.19 &     109 &  90 \\  
MT~259 & 4622.597 & Kcon(209)   & 0.34 & \phn 75 &  70 \\  
MT~299 & 4748.240*& Jcon(121)   & 0.08 & \phn 76 &  90 \\  
       & 4797.187 & Jcon(121)   & 0.08 & \phn 76 &  90 \\  
       & 4603.627 & Kcon(209)   & 0.35 & \phn 83 &  90 \\ 
MT~304 & 3623.314 & Jcon(112)   & 0.08 & \phn 84 &  90 \\  
       & 4801.189 & Kcon(209)   & 0.26 & \phn 75 &  90 \\  
MT~317 & 4607.626 & Kcon(209)   & 0.28 & \phn 85 &  90 \\  
MT~339 & 4610.621 & Kcon(209)   & 0.34 & \phn 84 &  90 \\  
MT~376 & 4747.386 & Jcon(121)   & 0.08 & \phn 82 &  90 \\  
       & 4610.632 & Kcon(209)   & 0.34 & \phn 81 &  60 \\  
       & 4612.522*& Kcon(209)   & 0.33 & \phn 85 &  48 \\  
MT~390 & 4608.512 & Kcon(209)   & 0.21 &     107 &  90 \\  
MT~403 & 4748.187 & Jcon(121)   & 0.07 & \phn 98 &  90 \\  
       & 4612.529 & Kcon(209)   & 0.24 &     102 &  90 \\  
MT~417 & 3613.365 & Jcon(112)   & 0.05 &     113 &  90 \\  
       & 3613.378 & Hcon(157)   & 0.12 & \phn 98 &  90 \\  
       & 3613.389 & Kcon(209)   & 0.28 & \phn 93 &  90 \\  
MT~421 & 5018.929 & J PHARO     & 0.04 &     141 &  50 \\  
       & 5018.926 & H PHARO     & 0.06 &     135 &  50 \\  
       & 5018.923 & K$_S$ PHARO & 0.06 &     134 &  50 \\  
       & 4740.217 & Kcon(209)   & 0.22 &     107 &  90 \\
MT~429 & 4748.202 & Jcon(121)   & 0.04 &     137 &  90 \\
       & 4622.616 & Kcon(209)   & 0.18 &     108 &  90 \\
MT~431 & 3625.266 & Jcon(112)   & 0.08 & \phn 87 &  90 \\
       & 3625.277 & Hcon(157)   & 0.19 & \phn 74 &  90 \\
       & 3625.288 & Kcon(209)   & 0.34 & \phn 79 &  90 \\
MT~448 & 4748.221 & Jcon(121)   & 0.04 &     112 &  90 \\
       & 4604.632 & Kcon(209)   & 0.34 & \phn 85 &  90 \\
MT~455 & 4752.240 & Jcon(121)   & 0.06 & \phn 93 &  90 \\
       & 4608.522 & Kcon(209)   & 0.28 & \phn 93 &  90 \\
MT~457 & 3613.401 & Jcon(112)   & 0.04 &     113 &  89 \\
       & 3613.414 & Hcon(157)   & 0.13 & \phn 82 &  95 \\
       & 3613.428 & Kcon(209)   & 0.19 &     107 &  90 \\
MT~462 & 4752.254 & Jcon(121)   & 0.05 & \phn 98 & 110 \\
       & 4609.614 & Kcon(209)   & 0.27 & \phn 95 &  94 \\
MT~465 & 3622.246 & Jcon(112)   & 0.03 &     124 &  90 \\
       & 3622.262 & Hcon(157)   & 0.10 &     101 &  86 \\
       & 3622.272 & Kcon(209)   & 0.18 &     105 &  90 \\
MT~470 & 4748.313 & Jcon(121)   & 0.05 & \phn 99 &  86 \\
       & 4752.267*& Jcon(121)   & 0.05 & \phn 99 &  90 \\
       & 4624.592 & Kcon(209)   & 0.16 &     106 &  90 \\
MT~473 & 4798.187 & Jcon(121)   & 0.04 &     100 &  90 \\
       & 4611.592 & Kcon(209)   & 0.24 & \phn 99 &  90 \\
MT~480 & 4611.602 & Kcon(209)   & 0.22 &     106 &  90 \\
MT~483 & 3625.369 & Jcon(112)   & 0.05 &     108 &  90 \\
       & 3625.381 & Hcon(157)   & 0.15 & \phn 89 &  80 \\
       & 3625.393 & Kcon(209)   & 0.23 &     104 &  90 \\
MT~485 & 4611.610 & Kcon(209)   & 0.24 &     101 &  90 \\
MT~507 & 4624.605 & Kcon(209)   & 0.17 &     108 &  90 \\
MT~516 & 3632.362 & Jcon(112)   & 0.02 &     151 &  90 \\
       & 3632.380 & Hcon(157)   & 0.07 & \phn 82 &  90 \\
       & 3632.390 & Kcon(209)   & 0.15 & \phn 81 &  90 \\
MT~531 & 4752.286 & Jcon(121)   & 0.05 & \phn 98 &  99 \\
       & 4612.619 & Kcon(209)   & 0.19 &     108 &  91 \\
MT~534 & 4605.612 & Kcon(209)   & 0.19 &     110 &  90 \\
MT~555 & 4613.606 & Kcon(209)   & 0.19 &     109 &  90 \\
MT~556 & 4752.223 & Jcon(121)   & 0.05 &     100 & 111 \\
       & 4613.616 & Kcon(209)   & 0.21 &     106 &  90 \\
MT~588 & 4594.622 & Kcon(209)   & 0.25 & \phn 97 &  90 \\
MT~601 & 4752.194 & Jcon(121)   & 0.05 &     103 &  90 \\
       & 4607.635 & Kcon(209)   & 0.22 &     104 &  90 \\
MT~605 & 4752.206 & Jcon(121)   & 0.03 &     151 &  90 \\
       & 4613.626 & Kcon(209)   & 0.15 &     148 &  90 \\
MT~611 & 4817.194 & Jcon(121)   & 0.04 &     111 &  90 \\
       & 4605.623 & Kcon(209)   & 0.22 &     102 &  90 \\
MT~632 & 4751.184 & Jcon(121)   & 0.02 &     119 &  90 \\
       & 4614.548 & Kcon(209)   & 0.18 &     106 &  90 \\
MT~642 & 4753.220 & Jcon(121)   & 0.04 &     107 &  90 \\
       & 4617.583 & Kcon(209)   & 0.21 &     103 &  90 \\
MT~692 & 4618.600 & Kcon(209)   & 0.17 &     112 &  78 \\
MT~696 & 4617.594 & Kcon(209)   & 0.19 &     106 &  90 \\
MT~716 & 4740.204 & Kcon(209)   & 0.22 &     107 &  90 \\
MT~734 & 4753.231 & Jcon(121)   & 0.04 &     109 &  90 \\
       & 4614.621 & Kcon(209)   & 0.43 & \phn 82 &  17 \\
       & 4618.510*& Kcon(209)   & 0.22 &     103 &  63 \\
MT~736 & 4753.188 & Jcon(121)   & 0.04 &     109 &  90 \\
       & 4627.467 & Kcon(209)   & 0.21 &     101 &  90 \\
MT~745 & 4607.554 & Kcon(209)   & 0.20 &     104 &  90 \\
MT~771 & 4753.240 & Jcon(121)   & 0.04 &     107 &  90 \\
       & 4607.564 & Kcon(209)   & 0.20 &     106 &  90 \\
MT~793 & 4753.206 & Jcon(121)   & 0.05 & \phn 95 &  90 \\
       & 4617.606 & Kcon(209)   & 0.24 & \phn 98 &  91 \\
S~3    & 3626.319 & Jcon(112)   & 0.03 &     124 &  87 \\
       & 3626.361 & Hcon(157)   & 0.07 &     117 &  90 \\
       & 3626.370 & Kcon(209)   & 0.15 &     115 &  90 \\
S~5    & 4593.620 & Kcon(209)   & 0.28 & \phn 87 &  90 \\
S~73   & 4593.629 & Kcon(209)   & 0.19 &     101 &  90 \\
WR~145 & 4740.194 & Kcon(209)   & 0.26 & \phn 97 &  90 \\
\enddata
\tablerefs{$^{C}$ Denotes that the combined image from both 
nights was used for analysis.\\
* Denotes which individual night was used for analysis.}
\end{deluxetable}

\clearpage

The NIRI data were reduced using the tools provided as part of the
Gemini reduction package in IRAF.  With the images rotated, reduced,
and the data quality robustly quantified through the various reduction
steps, we used two different combining programs to co-add all of the
frames. Most of the images were co-added using the IRAF tool IMCOADD
to derive an average image taking into account the bad pixel mask. In
the cases where IMCOADD failed (i.e., poor seeing, observations over
multiple nights, or blended point spread functions), GEMCOMBINE was
used with manual input of the central star pixel position. GEMCOMBINE
produces a slightly different median image than the mean coadded
IMCOADD, but the capability of allowing the user to define the pixel
shifts makes the final co-added image better aligned than when IMCOADD
fails. The final images from GEMCOMBINE and IMCOADD produce a slightly
larger field of view than the $22\arcsec \times 22\arcsec$ FOV of a
single frame, but depending on the observing conditions (e.g.,
exposure time and observations spanning multiple nights) some fields
can be larger than others. The PHARO data were reduced by debiasing,
flat fielding, bad pixel correction, and background subtraction and
then shift-and-added to create a single image.

We identified possible point sources by visually inspecting each frame
using SAO Image display software. This proved more successful than
automated methods due to the abundance of hot pixels from the IR
detector confused as point sources.  The faintest companions that we 
detect ($\Delta K \approx 9$~mag) have signals that are just above 
the threshold set by the readnoise of the camera and the number of 
coadded frames. We identified at least one source
in addition to the main target in each $K$-band frame through visual
inspection.  After identifying each point source and estimating the
approximate pixel position of its peak, we used SExtractor
\citep{ber96} to find each source and measure the centroid position
and relative brightness. The positions were determined from the {\tt
  XWIN\_IMAGE} and the {\tt YWIN\_IMAGE} keywords in SExtractor. The
relative flux returned by SExtractor is measured using the {\tt
  FLUX\_APER} parameter, which estimates the flux above the background
within a circular aperture.  We used nine aperture diameters on each
star to create an enclosed energy curve.  For close systems with
blended point spread functions (PSFs) ($\rho \leq 0\farcs1$), we used
a PSF deconvolution program, FITSTARS \citep{ten96,ten00}, to measure
the differential magnitude and separation.


\section{Results}\label{niriresults}     

We present the astrometric and photometric
results for all the stars in Table~\ref{t4-niri}
(given in full in the electronic version). 
The relative magnitudes and positions are determined with
respect to the target stars.   The columns of Table~\ref{t4-niri}
give the main target name, the angular separation $\rho$ and 
position angle $\theta$ (measured east from north) of the companion, 
its celestial coordinates, the magnitude difference and 
uncertainty in $J$, $H$, and $K$, the probability of chance 
alignment with a background field star $P_{ca}$ (see section 4.2), 
the identification number in the UKIRT Infrared Deep Sky Survey (UKIDSS)
\citep{law07}, and notes indicating other names, correspondence 
in another field, or measurement by FITSTARS (FS).   
The first row for a given target corresponds to the bright 
central star, and succeeding rows list data where available for
each detected companion star (arranged in order of increasing 
separation).  


\begin{longrotatetable}
\begin{deluxetable}{lrrccrrcccl}
\tablewidth{0pc}
\tabletypesize{\scriptsize}
\setlength{\tabcolsep}{0.1in} 
\rotate
\tablenum{4}
\tablecaption{Stars Detected in Sample \label{t4-niri}}
\tablewidth{0pt}
\tablehead{
\colhead{Field} &
\colhead{$\rho$} &
\colhead{$\theta$} &
\colhead{R.A.} &
\colhead{Dec.} &
\colhead{$\Delta J$} &
\colhead{$\Delta H$} &
\colhead{$\Delta K$} &
\colhead{$P_{ca}$} &
\colhead{UKIDSS} &
\colhead{Notes} \\
\colhead{Name} &
\colhead{(arcsec)} &
\colhead{(deg)} &
\colhead{(HH:MM:SS)} &
\colhead{(DD:MM:SS)} &
\colhead{(mag)} &
\colhead{(mag)} &
\colhead{(mag)} &
\colhead{} &
\colhead{Number} &
\colhead{} }
\startdata
A 11   & \nodata & \nodata & 20:32:31.543 & +41:14:08.21 & \nodata         & \nodata         & \nodata         & \nodata & 438717749790 &  MT 267                   \\
       &    0.77 &   282.6 & 20:32:31.476 & +41:14:08.38 & \nodata         & \nodata         &   4.71$\pm$0.03 &  0.000  & \nodata      &                           \\
       &    1.28 &   175.6 & 20:32:31.552 & +41:14:06.93 & \nodata         & \nodata         &   7.68$\pm$0.13 &  0.005  & \nodata      &                           \\
       &    2.20 &   276.9 & 20:32:31.350 & +41:14:08.48 & \nodata         & \nodata         &   4.17$\pm$0.02 &  0.002  & 438717693534 &                           \\
       &    3.66 &   103.8 & 20:32:31.858 & +41:14:07.34 & \nodata         & \nodata         &   9.00$\pm$0.31 &  0.068  & \nodata      &                           \\
       &    5.26 &   195.1 & 20:32:31.422 & +41:14:03.13 & \nodata         & \nodata         &   6.15$\pm$0.06 &  0.037  & 438717693527 &                           \\
       &    5.89 &   179.4 & 20:32:31.549 & +41:14:02.33 & \nodata         & \nodata         &   8.63$\pm$0.21 &  0.141  & 438717712064 &                           \\
A 12   & \nodata & \nodata & 20:33:38.217 & +40:41:06.40 & \nodata         & \nodata         & \nodata         & \nodata & 438262710179 &                           \\
\enddata
\end{deluxetable}
\end{longrotatetable}
\clearpage

\subsection{Astrometry}

The calibration of the astrometric transformation from pixel position 
of the PSF peak to the coordinates of the star is described in Appendix A. 
The celestial coordinates reported in Table \ref{t4-niri} 
are based upon the 2MASS coordinates of the primaries \citep{skr06}
and the relative SExtractor positions from the NIRI $K$-band images. 
We caution that in a few cases where a bright close companion exists, 
the 2MASS position refers to the center of light of the flux blend,
so the coordinates for all the associated companions target may have 
small systematic offsets in such cases.  However, the relative  
coordinate offsets from the main target derived from $(\rho, \theta)$
are reliable even in these cases.
The astrometry information listed for MT~421 includes a few
stars that were only observed with the PHARO camera, 
and for those the position on the Palomar $K_S$-band frame is used.

The uncertainties in the separation $\rho$ depend primarily on the 
pixel scale (known within $0.1\%$), non-linearity in the pixel scale 
(increasing uncertainty with separation), and uncertainties in 
centroid fitting where the PSFs of close pairs overlap.  
In general, the uncertainty in $\rho$ is less than $0\farcs07$. 
The position angle has a systematic uncertainty of $0\fdg1$ 
and a measurement uncertainty that is inversely proportional to $\rho$
(generally less than $0\fdg6 / \rho(\arcsec)$ in the absence of pair blending). 

\subsection{Photometry}\label{photo}

Most of the companions detected have separations $\rho > 0\farcs5$, 
and for these we relied upon the aperture photometry from SExtractor. 
We describe in Appendix B how the differential photometry calibration 
is accomplished by construction of enclosed energy curves for nine 
apertures of successively larger diameter for each detected star. 
We select from these measurements the aperture result with the largest
S/N ratio and then apply an appropriate aperture correction. 
The aperture correction is based upon the radial distance of the 
star from the center of the FOV and the seeing at the time of 
the observation, so that a first order correction may be made 
for the PSF degradation (lower Strehl ratio) with increasing off axis 
position.   Stars detected near the periphery of the FOV were 
measured in specially constructed edge-images that were formed 
from a subset of observations with optimized dither positions. 
Note that in the case of MT~421 the photometry is derived from the 
PHARO camera alone, because the NIRI results were limited to the $K$-band. 

There are also close systems with separations
$\rho < 0\farcs5$, where the companion falls within the halo
of the primary's PSF.  There are a total of nine such 
systems in our sample: A~20, A~26, A~41, MT~5, MT~429,
MT~605, MT~632, MT~642, and Schulte~73.
The PSFs are too blended for these close systems to use the 
aperture photometry performed by SExtractor. Instead, the 
photometric measurements were made using the program
FITSTARS, a PSF deconvolution program \citep{ten96,ten00}.  
FITSTARS fits blended PSFs to estimate the relative magnitudes and 
positions of the two components.  The code begins with a PSF 
estimate from an image of single star, and then uses an iterative
scheme to improve the specific PSF shape based upon the image 
of the binary star.  The outer wings of the PSF are constrained 
to be spherically symmetric.  The positions and amplitudes 
of the PSF for each component are optimized to minimize the 
residuals between the observations and model fit.  Numerical
tests with artificial companions are used to estimate the 
uncertainties in relative position and intensity.  Visual 
inspection of the residuals indicated that a simple two-component
fit was adequate in each of the nine cases where FITSTARS was applied.


\section{Detection of Physical Companions}\label{niridisc}   

\subsection{Detection Limits}

\begin{figure}
\begin{center}
{\includegraphics[angle=90,width=15cm]{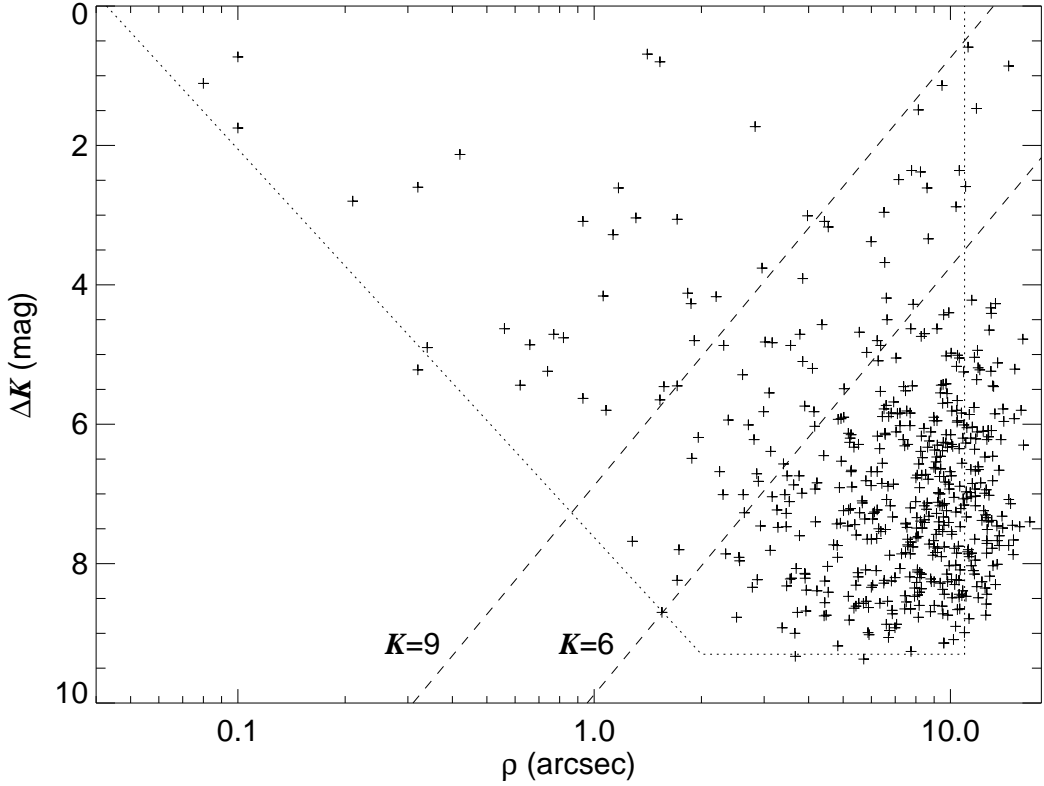}}
\end{center}
\caption{The detected companions as a function of angular separation $\rho$
and magnitude difference $\Delta K$.  The dotted lines show the 
approximate lower limit for positive detection within our sample. 
The two diagonal dashed lines indicate the lower limits for 
meeting the chance alignment with background stars criterion $P_{ca}<1\%$ 
for primary star magnitude $K=6$ and 9 mag.  Only those companions
above both the dotted and dashed lines are included in the 
assessment of binary statistics.  \label{f1-niri}}
\end{figure}

We made one epoch imaging of 74 O- and B-type stars in Cyg OB2 with
high angular resolution methods in the infrared $JHK$ bands, and we
found at least one star in the field around each of our targets, for a
total of 546 possible companions.  We present in Table~\ref{t4-niri}
photometric and positional information for stars found in the field
around our targets.  Figure \ref{f1-niri} shows the dynamical range of
our detections as a function of separation. This figure demonstrates
the sensitivity and completeness of our survey. The separation axis is
plotted as $\log\rho$ to show the sensitivity at both close and large
distances.  The closest pair resolved was the binary MT~429 with $\rho
= 0\farcs08$ (3.6 pixels), while the largest separation was $\rho =
16\farcs71$ for a distant star in the FOV of MT~421.  The relatively
faintest companion (of MT~716) has a magnitude difference of $\Delta K
= 9.37$~mag.  The dotted lines in Figure \ref{f1-niri} show the
approximate limits for detection in our sample that are bounded by the
largest contrast ratio at the bottom, half of the square FOV on the
right, and the restriction to brighter companions at closer separation
on the left.  The limiting dotted line in Figure~1 is substantially the same as the 
detection limit found by \citet{laf14} (see their Fig.\ 1) who calculated 
the standard deviation as a function of separation in annuli of 
residual images with companions removed.  The work by \citet{laf14} is
based upon the same NIRI/ALTAIR camera system as we used, and their 
target sample spans a similar magnitude range, and thus their detection 
threshold is essentially the same as we plot in Figure~1.
Note that the exposure times were selected to obtain a uniform S/N 
ratio for all the targets, so the detection limits are generally 
the same for bright and faint targets (Table 3 documents the relatively
small differences in image quality and Strehl ratio between observations).
The faint limit shown in Figure~1 applies generally to parts of the co-added 
image with $\rho < 10\arcsec$, and detection limit is degraded in the 
outer parts where the sky is only recorded in a subset of the dither 
positions.

We performed several numerical experiments where we
created artificial binaries to test the detection limits. The lower
limit of magnitude as a function of separation was similar to the area
bounded by the dotted lines in Figure \ref{f1-niri}.


\subsection{Probable Bound Companions}

\begin{figure}
\begin{center}
{\includegraphics[angle=90,width=15cm]{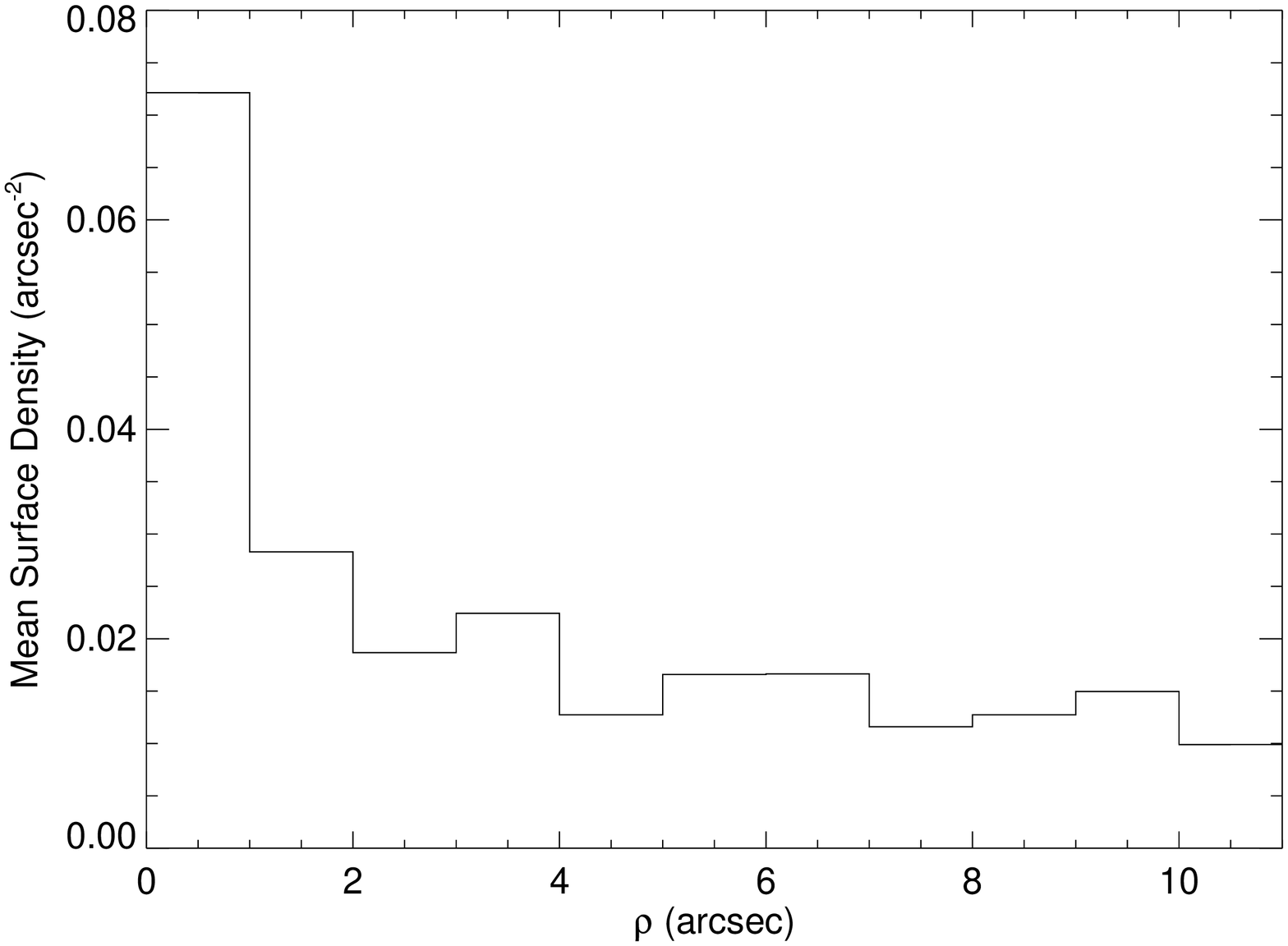}}
\end{center}
\caption{The surface density of stellar companions as a
function of angular separation $\rho$.  The peak at small $\rho$
probably corresponds to physical companions, while the numbers
at larger distance reflect more the typical star count background in
the direction of Cyg~OB2.  At a distance of $1.33$~kpc, $1\arcsec$
corresponds to a projected separation of 1330~AU. \label{f2-niri}}
\end{figure}

Ideally we would like to differentiate between chance alignments and
gravitationally bound systems. The best way to do so is to obtain
multi-epoch observations, in conjunction with a proper motion
study and spectroscopic information about the companions.
However, for this study we have only a single epoch observation and
$JHK$ color information.  In Figure \ref{f2-niri}, we show the number
density (number arcsec$^{-2}$) of companions for the entire sample as 
a function of separation. The companion density levels off at a
separation of $\rho \approx 4\arcsec$.  This very likely corresponds
to the average number density of stars in the association and along
this line of sight.  Stars found at separations $\rho > 4\arcsec$
are more likely to be chance alignments. However, the surface density
increases greatly within $\rho < 1\arcsec$, and stars found within
this separation are more likely to be physically bound companions.


Since we only have access to single-epoch observations,
we may apply a statistical argument developed by \citet{cor06}
to determine likely companions.   The statistical probability
that a detected companion is part of the background field of
stars in this direction was calculated using the expression from 
\citet{cor06},
\begin{equation}
P_{ca} ( \Sigma_{K},\rho ) = 1 - e^{-\pi\Sigma_{K}{\rho}^{2}}.
\end{equation}
Here $P_{ca}$ is the probability of finding a field star within 
a circle with a radius $\rho$ (in arcseconds) centered on the target
(subscript ``ca'' refers to chance alignment). 
$\Sigma_{K}$ is the cumulative surface density of stars (arcsec$^{-2}$) 
in the surrounding field that includes all stars brighter than 
magnitude $K$.  Our working assumption is that if the probability
$P_{ca}$ is low, then the detected companion is likely to be 
physically associated with the target.  

The field surface density was determined using
a combination of data from 2MASS \citep{skr06} and UKIDSS
\citep{law07} of the area surrounding around each of our
targets. The 2MASS survey provided photometry for stars with $K <
14$~mag and UKIDSS provided the information for $14< K < 16$~mag.
The magnitudes of faint stars in UKIDSS were set by comparing the
magnitudes of stars in the range of $K=8-14$~mag where the 
2MASS and UKIDSS sets of observations overlapped. 
This was done by identifying stars in the
UKIDSS frame that had 2MASS $K$ magnitudes. Then the magnitudes of
fainter stars were determined from the stars in common with 2MASS and
the differential magnitudes measured in the UKIDSS catalog.
We formed areal density estimates $\Sigma_{K}$ in bins of one magnitude
increments for tabular interpolation purposes.  The field star counts
increase rapidly towards fainter magnitudes, and an approximate linear 
fit of the mean star count trend is $\log \Sigma_{K} = -7.67 +0.326 K$.
The binned version of $\log \Sigma_{K}$ is in good agreement with 
predicted star counts for the direction of Cyg~OB2 from the 
Besan{\c{c}}on model of the Galaxy\footnote{https://model.obs-besancon.fr}
\citep{cze14} over the range of $K=7$ to 15 mag. 

The derived cumulative star count function $\Sigma_{K}$ is based 
on all the stars in the Cyg~OB2 fields including the targets and 
any physical companions.  Consequently, $\Sigma_{K}$ may overestimate
the numbers of field association, foreground, and background stars in 
the vicinity of the targets, because the physical companions are 
included.  The result is that the probability of finding a field 
star $P_{ca}$ increases, so that some physical companions may be 
placed in the field rather than bound categories.  Thus, we may 
be rejecting some physical companions from consideration, 
especially at the brighter end where the targets and their 
bound companions contribute most to the net star counts.  
However, this potential underestimate of the numbers of 
physical companions is negligible, because the stars in Cyg~OB2 
are dispersed over a large part of the sky \citep{wri15} and 
the areal density of bound companions is low.  The good match 
of our empirical $\Sigma_{K}$ relation to the Galactic model 
for background stars confirms that the relative contribution 
of Cyg~OB2 stars to the total star counts is low, especially 
towards fainter stars.  Ideally we might consider a star count
model that includes components from both the field and 
bound companions, but the latter would need many apriori 
assumptions about the number distributions of physical companion 
mass and separation that are poorly known at present. 

We estimated the probability $P_{ca}(\Sigma_{K},\rho)$ based 
upon the companion magnitude $K$ and separation $\rho$. 
The $K$ magnitudes of the companions were determined from 
the 2MASS $K_S$ magnitude of the primary plus the $\Delta K$ 
magnitude from the NIRI observations.  Then the predicted
field star areal density was estimated by linear interpolation 
in the $(K, \log \Sigma_{K})$ plane (and by extrapolation for 
the faintest companions).  Finally, we used the functional 
expression for $P_{ca}(\Sigma_{K},\rho)$ given above to estimate 
the field star chance alignment probability for each detected companion. 
The calculated probability $P_{ca}(\Sigma_{K},\rho)$ is listed
in column 9 of Table \ref{t4-niri}.   We assume
that the companions with $P_{ca}(\Sigma_{K},\rho$)$ < 1\%$
are unlikely to be members of the field population, and are 
instead physical companions located near to their respective 
target star.  The numbers of such probable physical companions 
are summarized in Table \ref{t5-niri} (given in full in
the electronic version).  The columns give the
star name, total number of stars in the NIRI FOV, the number
of probable companions, the number of companions found in the
{\it HST}/FGS high angular resolution survey by \citet{cab14},
the number of close companions found as spectroscopic binaries
by \citet{kob14}, the total number of all known companions
(astrometric and spectroscopic), the number of companions new
to this work, and the mass of the central star based upon its
position in the H-R diagram from \citet{wri15}.  The companions
detected in the {\it HST}/FGS survey are all confirmed here with the
exception of the very close companions of MT~304 ($\rho = 0\farcs064$)
and MT~696 ($\rho = 0\farcs023$) that are too close and faint
for resolution with NIRI.   On the other hand, the NIRI imaging
program has revealed fainter companions that eluded detection with FGS.
There are 25 new detections in our NIRI survey that were
unknown companions prior to this work.


\startlongtable
\begin{deluxetable}{lccccccc}
\tablewidth{0pc}
\tabletypesize{\scriptsize}
\tablenum{5}
\tablecaption{Multiplicity Properties\label{t5-niri}}
\tablewidth{0pt}
\tablehead{
\colhead{Star} &
\colhead{$N$(FOV)} &
\colhead{$N(P_{ca}< 0.01$)} &
\colhead{$N$(FGS)} &
\colhead{$N$(SB)} &
\colhead{$N$(Total)} &
\colhead{$N$(New)} &
\colhead{$M_1/M_\odot$} \\
\colhead{(1)} &
\colhead{(2)} &
\colhead{(3)} &
\colhead{(4)} &
\colhead{(5)} &
\colhead{(6)} &
\colhead{(7)} &
\colhead{(8)} }
\startdata
A~11   & 7  & 3       & \nodata & 1       & 4       & 3  &    34.7     \\
A~12   & 3  & 0       & \nodata & \nodata & 0       & 0  &    \nodata  \\
A~15   & 4  & 0       & \nodata & \nodata & 0       & 0  &    31.8     \\
A~18   & 12 & 0       & \nodata & \nodata & 0       & 0  &    \nodata  \\
A~20   & 6  & 1       & \nodata & \nodata & 1       & 1  &    35.0     \\
A~23   & 2  & 0       & 0       & \nodata & 0       & 0  &    26.3     \\
A~24   & 3  & 0       & \nodata & \nodata & 0       & 0  &    29.6     \\
A~25   & 7  & 0       & \nodata & \nodata & 0       & 0  &    \nodata  \\
A~26   & 7  & 1       & \nodata & \nodata & 1       & 1  &    18.7     \\
A~27   & 3  & 0       & 0       & \nodata & 0       & 0  &    35.2     \\
A~29   & 2  & 0       & \nodata & \nodata & 0       & 0  &    \nodata  \\
A~32   & 7  & 0       & \nodata & \nodata & 0       & 0  &    \nodata  \\
A~37   & 6  & 0       & \nodata & \nodata & 0       & 0  &    \nodata  \\
A~38   & 12 & 1       & \nodata & \nodata & 1       & 1  &    20.3     \\
A~41   & 3  & 1       & 1       & \nodata & 1       & 0  &    \nodata  \\
A~46   & 4  & 0       & 0       & \nodata & 0       & 0  &    \nodata  \\
B~17   & 5  & 0       & \nodata & 1       & 1       & 0  &    24.8     \\
MT~5   & 16 & 1       & 1       & 0       & 1       & 0  &    \nodata  \\
MT~59  & 3  & 1       & 1       & 1       & 2       & 0  &    25.5     \\
MT~70  & 9  & 0       & 0       & 1       & 1       & 0  &    18.4     \\
MT~83  & 3  & 0       & 0       & 0       & 0       & 0  &    14.3     \\
MT~138 & 9  & 1       & 1       & 0       & 1       & 0  &    23.6     \\
MT~140 & 9  & 0       & \nodata & \nodata & \nodata & \nodata  & \nodata \\
MT~145 & 14 & 0       & 0       & 1       & 1       & 0  &    16.8     \\
MT~213 & 7  & 0       & 0       & 0       & 0       & 0  &    14.5     \\
MT~217 & 6  & 2       & 0       & 0       & 2       & 1  &    28.6     \\
MT~227 & 7  & 0       & 0       & 0       & 0       & 0  &    19.1     \\
MT~250 & 21 & 0       & 0       & 0       & 0       & 0  &     8.0     \\
MT~258 & 16 & 1       & 0       & 1       & 2       & 1  &    21.9     \\
MT~259 & 16 & 0       & 0       & 0       & 0       & 0  &    12.9     \\
MT~299 & 7  & 1       & 0       & 0       & 1       & 1  &    23.4     \\
MT~304 & 2  & 1       & 1       & 0       & 2       & 0  &   110.0     \\
MT~317 & 4  & 0       & 0       & 0       & 0       & 0  &    24.8     \\
MT~339 & 11 & 0       & 0       & 1       & 1       & 0  &    21.2     \\
MT~376 & 8  & 0       & 0       & 0       & 0       & 0  &    20.9     \\
MT~390 & 10 & 0       & 0       & 1       & 1       & 0  &    23.5     \\
MT~403 & 11 & 1       & 0       & 1       & 2       & 1  &    10.7     \\
MT~417 & 12 & 2       & 2       & 1       & 3       & 0  &    49.9     \\
MT~421 & 47 & 4       & \nodata & 1       & 5       & 1  &    16.3     \\
MT~429 & 9  & 1       & 1       & 2       & 3       & 0  &    13.5     \\
MT~431 & 2  & 0       & 0       & 1       & 1       & 0  &    51.6     \\
MT~448 & 17 & 0       & 0       & 1       & 1       & 0  &    28.6     \\
MT~455 & 19 & 0       & 0       & 0       & 0       & 0  &    21.4     \\
MT~457 & 2  & 0       & 0       & 0       & 0       & 0  &    46.7     \\
MT~462 & 12 & 3       & 0       & 0       & 3       & 1  &    35.2     \\
MT~465 & 6  & 3       & 0       & 1       & 4       & 0  &    41.1     \\
MT~470 & 12 & 1       & 0       & 0       & 1       & 1  &    16.8     \\
MT~473 & 16 & 0       & 0       & 2       & 2       & 0  &    20.2     \\
MT~480 & 11 & 0       & 0       & 0       & 0       & 0  &    25.4     \\
MT~483 & 4  & 1       & 0       & 0       & 1       & 1  &    41.6     \\
MT~485 & 10 & 0       & 0       & 1       & 1       & 0  &    21.8     \\
MT~507 & 7  & 0       & 0       & 0       & 0       & 0  &    18.7     \\
MT~516 & 4  & 2       & 1       & 0       & 2       & 1  &    51.6     \\
MT~531 & 6  & 1       & 1       & 0       & 1       & 0  &    23.5     \\
MT~534 & 6  & 0       & 0       & 0       & 0       & 0  &    23.4     \\
MT~555 & 4  & 0       & 0       & 1       & 1       & 0  &    24.9     \\
MT~556 & 6  & 1       & 0       & 0       & 1       & 1  &    28.9     \\
MT~588 & 11 & 0       & 0       & 1       & 1       & 0  &    17.9     \\
MT~601 & 12 & 1       & 0       & 1       & 2       & 1  &    26.0     \\
MT~605 & 9  & 1       & 1       & 1       & 2       & 0  &    12.5     \\
MT~611 & 9  & 1       & 0       & 0       & 1       & 1  &    22.3     \\
MT~632 & 4  & 3       & 1       & 0       & 3       & 2  &    37.4     \\
MT~642 & 5  & 1       & 0       & 0       & 1       & 1  &    15.9     \\
MT~692 & 10 & 0       & 0       & 0       & 0       & 0  &    14.2     \\
MT~696 & 8  & 1       & 1       & 1       & 3       & 1  &    17.2     \\
MT~716 & 17 & 1       & \nodata & 0       & 1       & 1  &    17.5     \\
MT~734 & 2  & 0       & 0       & 1       & 1       & 0  &    43.7     \\
MT~736 & 3  & 0       & 0       & 0       & 0       & 0  &    18.0     \\
MT~745 & 3  & 0       & 0       & 1       & 1       & 0  &    25.4     \\
MT~771 & 5  & 1       & 0       & 1       & 2       & 1  &    29.0     \\
MT~793 & 7  & 0       & 0       & 0       & 0       & 0  &    12.7     \\
S~3    & 4  & 1       & \nodata & 1       & 2       & 0  &    38.0     \\
S~5    & 3  & 2       & 1       & 2       & 4       & 0  &    93.1     \\
S~73   & 13 & 1       & 0       & 1       & 2       & 1  &    19.6     \\
WR~145 & 2  & 0       & 0       & 1       & 1       & 0  & $> 25  $    \\
\enddata
\tablecomments{\\
(1) Star Name \\
(2) Total number of the stars found in the target field.\\
(3) Number of high probability companion stars from NIRI.\\
(4) Number of companions detected with FGS.\\
(5) Number of companions found through radial velocity measurements
    (\citealt{kob14}, or for the case of WR~145, \citealt{mun09}, and
    references therein). \\
(6) Total number of unique companions from columns (3) through (5).\\
(7) New companions detected during this work.\\
(8) Mass of primary from \citet{wri15}.}
\end{deluxetable}
\clearpage

\subsection{Color-Magnitude Diagram of the Companions}

\begin{figure}[ht!]
\begin{center}
{\includegraphics[angle=90,width=15cm]{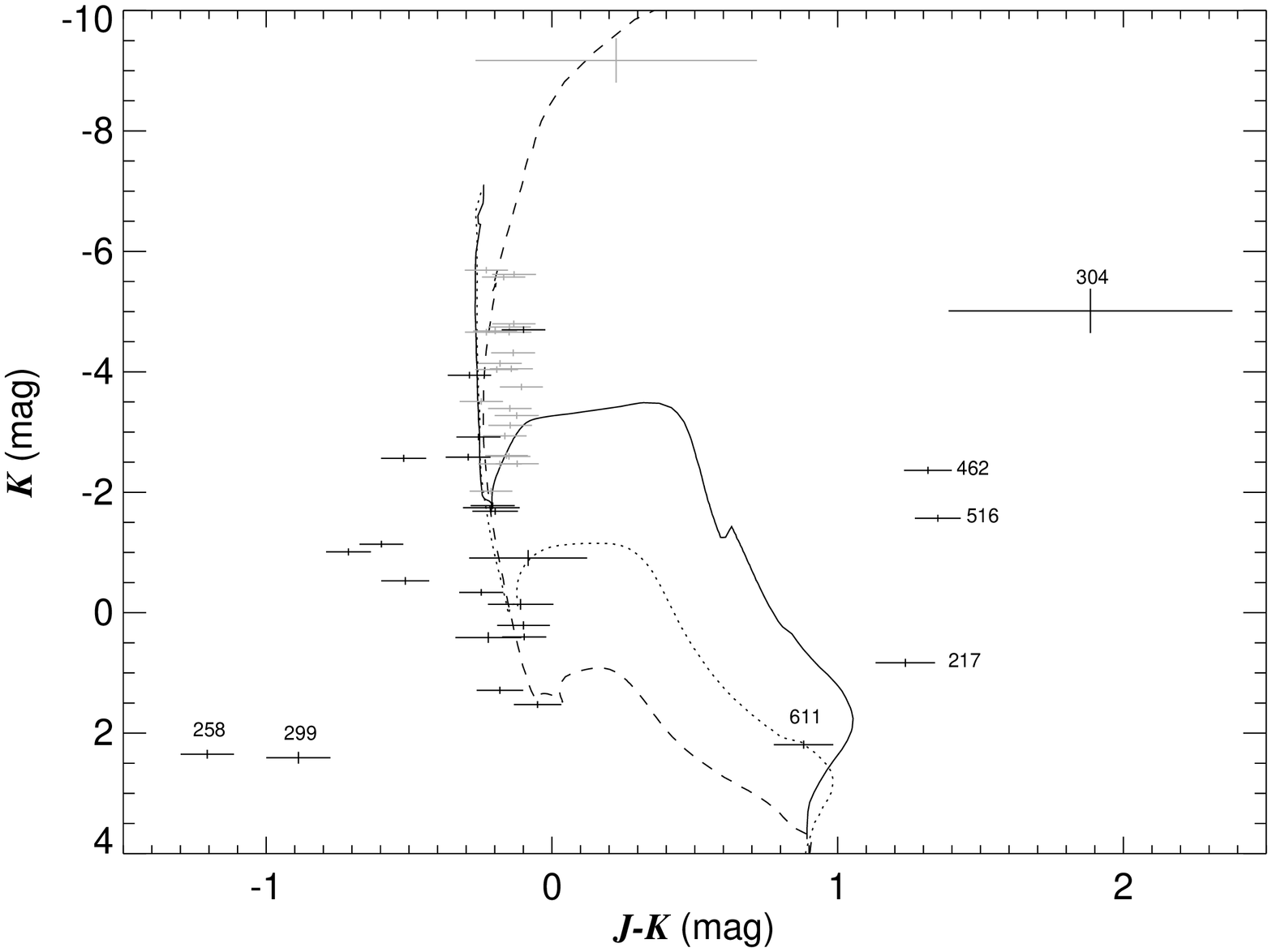}}
\end{center}
\caption{Color-magnitude diagram in $J-K_s$ and $K_s$ of the probable
  companions (black) and their primary stars (gray), dereddened
  according to estimates from \citet{neg08}, \citet{mas91}, and
  \citet{tor91}, and converted to absolute magnitude using a distance
  modulus of $DM =10.66$.  Overlaid are isochrones for ages of 0.1~Myr
  (solid line), 1~Myr (dotted line), and at 7~Myr (dashed line) from
  \citet{bre12}. Very blue and red companions are labeled by their
  primary's MT number.\label{f3-niri}}
\end{figure}

We can determine some facts about the nature of the probable
companions by plotting their positions in a near-IR color - magnitude
diagram $(J-K,K)$.  We constructed such a diagram for those targets
with probable companions in the following way.  We began by converting
the relative magnitudes $\Delta J$ and $\Delta K$ to actual magnitudes
by adding these to the 2MASS magnitudes $J$ and $K_S$ for the central
target.  In a few cases we needed to adjust the 2MASS magnitudes to
remove the flux of companions within $3\arcsec$ of the central star
that contributed to the total flux recorded in the lower angular
resolution 2MASS measurements.  Next we dereddened each of the $J$ and
$K$ magnitudes using the reddening associated with the primary target.
The reddening values were adopted from one of three sources, listed in
Table \ref{t1-niri}, in order of preference and availability:
\citet{neg08} for $E(J-K)$, \citet{mas91} for $E(B-V)$, and
\citet{tor91} for $E(b-y)$.  We applied the extinction correction
transformations from \citet{fit99} to convert the adopted reddening to
the total extinction in the infrared, $A_J$ and $A_K$. We adopted the
default value of total to selective extinction $R = 3.1$, which is
slightly larger than $R=2.9$ found by \citet{wri15}.  We then combined
the two measurements to create the dereddened color index $J-K$.  The
highest accuracy distance estimates for Cyg~OB2 come from interstellar
maser parallax measurements by \citet{ryg12} and from eclipsing binary
dimensions by \citet{kim15}, and we adopted the error weighted mean of
their results to arrive at a distance of 1.36~kpc (distance modulus =
10.66 mag).  We used this distance to transform the extinction
corrected $K$ magnitude to absolute $K$ magnitude.  The resulting
color - magnitude diagram appears in Figure \ref{f3-niri}, where the
central targets are plotted as gray symbols and the probable companions
as black symbols.  For the sake of clarity, we omitted several cases with
uncertainties in color in excess of 0.9 mag.  Also plotted in Figure
\ref{f3-niri} are theoretical isochrones for three ages from the
PARSEC code\footnote{http://stev.oapd.inaf.it/cgi-bin/cmd}
\citep{bre12}.

We see that most of the central targets are close to the nearly
vertical main sequence track, with the exception of the evolved star
MT~304 = Cyg~OB2~\#12 found near the top of the diagram.  Likewise,
most of the companions also appear as lower mass main sequence stars
with implied masses down to $2 M_\odot$.  There are a few interesting
outliers that deserve comment.  The companions of MT~258 and MT~299
appear in the very blue and faint part of the color - magnitude
diagram, and we suspect that these are less reddened foreground
objects rather than physical companions.  There are also five very red
companion stars that appear to be far from the main sequence.  These
may be cool field stars, companions that are more reddened than their
primary stars, or pre-main sequence stars.  Given the youth of Cyg~OB2
(1 -- 7 Myr; \citealt{wri15}), some of these companions may have
retained natal disks that would contribute to their long-wavelength
flux.


\section{Multiplicity}\label{nirimultiplicity}  

We can use the total number of probable physical companions 
(column 3 of Table \ref{t5-niri}) to determine the multiplicity 
properites of our sample of 74 massive stars in Cyg~OB2. 
There are 26 single ($35\% $), 27 binary ($36\% $), 
12 triple ($16\% $), and 9 higher multiplicity systems ($12\% $).
Note that we have essentially double counted the numbers
in two cases where companions occur in two adjacent and
overlapping fields, MT~213 + MT~217 (Schulte 4A,B) and
MT~462 + MT~465 (Schulte 8A,B).   The target with the
largest number of companions (4) is MT~421 (Schulte 50) that
resides at the center of a tentatively identified star
cluster \citep{bic03}, so it it possible that some of the
companions that met the $P_{ca}$ criterion are cluster
members that are not directly orbiting the central star.

The total multiplicity fraction $MF$ (number of targets with 
any companion divided by the number of targets) and
companion frequency $CF$ (total number of companions 
divided by the number of targets) are summarized in 
Table \ref{t6-niri} with uncertainties estimated as 
described by \citet{ald15}.  The columns list the 
sample, a reference code, the number of primary targets $N$, 
the range in physical separation of the detected companions
given as the logarithm (base 10) of separation in 
Astronomical Units $\log a$, the maximum magnitude 
difference of companions $\triangle m$(max), $MF$, and $CF$.  
The top section of Table \ref{t6-niri} lists our results and those 
of prior studies for companions that are angularly resolved. 
The lower section gives similar statistics by including 
closer systems discovered as spectroscopic binaries and/or 
eclipsing and ellipsoidal binaries (SB/E) in order to 
estimate the multiplicity properties over the full range 
of separation.  The combined resolved and SB/E companion 
numbers are listed in column 6 of Table \ref{t5-niri}, and the 
resulting statistics are shown in the first row of the lower section 
of Table \ref{t6-niri}.  The multiplicity fraction increases 
from $MF=0.46$ to 0.65 by adding the known closely separated 
binaries, and likewise the companion frequency increases 
from $CF=0.69$ to 1.11 with inclusion of the close systems. 
There are 29 known spectroscopic systems among our sample 
of stars (column 5 of Table \ref{t5-niri}), and resolved companions 
are more common in this subset ($MF=15/29=0.52$ and $CF=24/29=0.83$)
than among the full sample ($MF=0.46$ and $CF=0.69$). 


\begin{deluxetable}{lrrrrcc}
\tablewidth{0pc}
\tablenum{6}
\tablecaption{Frequency of Multiple Systems and Companion Frequency\label{t6-niri}}
\tablewidth{0pt}
\tablehead{
\colhead{Sample}   &
\colhead{Ref.}\tablenotemark{a}     &
\colhead{$N$}      &
\colhead{$\log a$} &
\colhead{$\triangle m$(max)} &
\colhead{$MF$}     &
\colhead{$CF$}  \\
\colhead{}      &
\colhead{}      &
\colhead{}      &
\colhead{(AU)}  &
\colhead{(mag)} &
\colhead{}      &
\colhead{}  
}
\startdata
\cutinhead{Resolved Companions}
Cyg OB2                        &  1 &  74 &    [2,4] &  9 &  $0.46 \pm 0.06$ & $0.69 \pm 0.11$ \\
Cep OB2/3                      &  2 & 148 &    [2,3] &  7 &   0.25           &  0.27           \\
Orion Trapezium                &  3 &  16 &    [0,3] &  5 &   0.69           &  1.38           \\
Young Stars in Upper Sco       &  4 &  91 &    [1,3] & 10 &  $0.27 \pm 0.05$ & $0.43 \pm 0.07$ \\
Sco OB2 B-type                 &  5 &  58 &    [0,1] &  3 &   0.26           &  0.26           \\
NGC 6611                       &  6 &  60 &    [2,3] &  6 &  $0.18 \pm 0.06$ & $0.18 \pm 0.06$ \\
Galactic OB in clusters/assoc. &  7 & 214 &    [1,3] &  5 &  $0.31 \pm 0.03$ & $0.34 \pm 0.04$ \\
Southern O-type                &  8 &  96 &    [0,4] &  8 &  $0.75 \pm 0.04$ &  1.5            \\
Massive YSOs                   &  9 &  32 &    [3,5] &  5 &  $0.31 \pm 0.08$ & $0.53 \pm 0.09$ \\
\cutinhead{Resolved + SB/E Companions}
Cyg OB2 (all)                  & 10 &  74 & [$-1$,4] &  9 &  $0.65 \pm 0.05$ & $1.11 \pm 0.13$ \\
Cyg OB2 ($M<25 M_\odot$)       & 10 &  38 & [$-1$,4] &  9 &  $0.66 \pm 0.08$ & $1.00 \pm 0.17$ \\
Cyg OB2 ($M>25 M_\odot$)       & 10 &  27 & [$-1$,4] &  9 &  $0.78 \pm 0.08$ & $1.56 \pm 0.23$ \\
Galactic OB in clusters/assoc. &  7 & 214 & [$-1$,4] &  5 &  $0.69 \pm 0.03$ & $1.67 \pm 0.17$ \\
Southern O-type                &  8 &  96 & [$-1$,4] &  8 &  $0.91 \pm 0.03$ &  2.1            \\
\enddata
\tablenotetext{a}{
1. This paper and \citet{cab14}; 
2. \citet{pet12}; 
3. \citet{gra18};
4. \citet{laf14};
5. \citet{riz13};
6. \citet{duc01}; 
7. \citet{ald15}; 
8. \citet{san14}; 
9. \citet{pom19};
10. This paper, \citet{cab14}, and \citet{kob14}.
}
\end{deluxetable}

It is important to bear in mind that our
reported statistics on angularly resolved binaries only
include those companions above both the dotted (detection limited)
and dashed lines (background limited) in Figure \ref{f1-niri},
so the $MF$ and $CF$ results in Table \ref{t6-niri} should be
regarded as lower limits because we miss systems outside of 
these limits.  In particular, there is a systematic bias against 
detection of close and faint companions (Fig.\ \ref{f1-niri}).
Consequently, it is very difficult to derive 
distributions of binary separation and mass ratio from our
results.  Furthermore, the magnitude-dependent characteristics 
of these limits may introduce some biases into our results, 
for example, with respect to stellar mass. 

There is a well-known trend for the multiplicity fraction 
to increase with stellar mass \citep{duc13,san14}, and 
it is worthwhile examining whether or not this trend 
exists within our sample of Cyg~OB2 stars.  We divided 
the stars with mass estimates (column 8 of Table \ref{t5-niri})
into those below and above $25 M_\odot$, and the statistics
for these groups are given for the combined resolved plus SB/E 
companion numbers in rows 2 and 3 of the lower section of 
Table \ref{t6-niri}.  We see that both $MF$ and $CF$ are 
larger in the higher mass group as expected for the 
trend of increasing multiplicity with stellar mass.  
We caution, however, that this mass dependence
is partially due to selections effects.  We show in Figure \ref{f1-niri} 
the dividing lines for meeting the $P_{ca} < 1\% $ criterion 
for target stars with bright and faint magnitudes.
These trends show that at larger separation $\rho$ the
$P_{ca} < 1\% $ criterion will reject more and more brighter
companions because of confusion with the background field.
At the fainter apparent magnitude of the lower mass stars
in our sample, the exclusion of candidate binaries becomes
even more severe, so we expect that the multiplicity
fraction will be lower for lower mass stars because of
the greater difficulty in distinguishing their companions 
from the background stars.  Consequently, the apparent 
increase in $MF$ and $CF$ with stellar mass in Table \ref{t6-niri}
is probably overestimated.  Indeed, the statistics for the 
high mass group are probably more representative of 
the actual numbers, because the selection limits are more
generous for the brighter, massive targets. 

The upper part of Table \ref{t6-niri} compares our multiplicity 
results with other earlier investigations from adaptive optics, 
Lucky Imaging, and interferometry.  All these samples consist 
of massive or very young stars, similar to the composition of 
our Cyg~OB2 sample.  However, each of these surveys is sensitive
to a particular range in angular separation and maximum magnitude
contrast (columns 4 and 5, respectively, in Table \ref{t6-niri}), 
and in general those studies that cover a broader range in 
separation and magnitude difference yield higher multiplicity frequencies.  
Our results for $MF$ and $CF$ fall well within the range of these earlier 
studies, and higher values are only found from recent VLTI interferometric 
studies of the nearby Orion Trapezium \citep{gra18} and southern O-type 
stars \citep{san14}, and these studies span a relatively large 
range in separation and contrast sensitivity. 

The lower part of Table \ref{t6-niri} compares the statistics
for the combined wide and close binary samples of Cyg~OB2 with 
those from two all-sky surveys.  Our results are broadly consistent 
with those from the {\it HST}/FGS survey of O-type stars by \citet{ald15}
for their subset of cluster and association members and with the 
VLTI/PIONIER and NACO/Sparse Aperture Masking survey of southern sky O-stars 
by \citet{san14}.  In particular, if we adopt the results from the
high mass group as the least affected by selection effects (see above), 
then our $MF$ and $CF$ results for Cyg~OB2 appear to be 
consistent with these other surveys.  Taken together, these studies
imply that the massive stars in clusters and associations have a 
very large multiplicity frequency compared to lower mass stars
\citep{duc13}. 

The high incidence of multiple systems among the more massive
stars indicates that the angular momentum of the natal cloud is 
preferentially transformed into orbital motion \citep{lar10}.
The processes involved in massive star formation are still 
the subject of active investigation \citep{ros20}.  
The turbulent core model envisions the collapse 
of a virial natal cloud that creates widely spaced binaries 
accompanied by a small number of low mass stars formed by cloud 
fragmentation \citep{ros19}.  The stellar cores are surrounded
by large disks, and disk fragmentation can lead to the 
formation of bound stellar companions \citep{kra06}.
Subsequent disk accretion onto these companions can lead to the 
formation of close binaries \citep{lun18,tok20} that have 
much smaller separations than those investigated here. 
Alternatively, the competitive accretion model \citep{bon06}
suggests that massive stars form by accretion onto a cluster 
of low mass seeds in the dense, central regions of the natal 
cloud.  These models tend to form star clusters where 
three-body encounters can create massive binaries over 
a wide range in separation \citep{wal19}.  Both the 
turbulent core and competitive accretion models predict 
an increased binary fraction among more massive stars, 
but with somewhat different distributions in separation 
and mass ratio \citep{pet12,gra18}. 

The subsequent dynamical interactions in small number 
clusters will generally lead to the formation a single, 
wide massive binary and the ejection of lower mass 
single stars \citep{gri18}.  Wide binaries with 
separations of 100 to 10,000 AU are large enough for 
frequent gravitational encounters to occur in dense 
environments, and the large numbers of such wide 
binaries in Cyg~OB2 indicates that they have survived 
potential disruptive encounters.  \citet{gri18} argue 
that star formation in Cyg~OB2 probably occurred in 
many well-separated locations in the natal cloud, so 
that close encounters with other cluster stars did
relatively little damage to these wide binaries.  
This conclusion is bolstered by the fact that the 
binary frequency found in massive Young Stellar Objects 
(representing the frequency at birth) is similar to 
the present day binary frequency in Cyg~OB2 (Table \ref{t6-niri})
even after several million years of dynamical evolution. 
 

\section{Conclusions}\label{niriconclusions}     

Our near-IR adaptive optics survey of the Cyg~OB2 association
has yielded astrometry and photometry for the fields surrounding 
74 of its massive O- and B-type members.  
We find that $46\% $ of the sample of stars
have a companion that is probably physically related.
These companions have projected separations in the 
range from 100 to 19,000 AU, and the faintest 
companions detected are probably $2 M_\odot$ stars
based upon their positions in the $(J-K, K)$ color -
magnitude diagram.  Many other closer companions must
exist, and we included spectroscopic binary results from 
studies by \citet{kob14} that primarily sample systems 
with a semimajor axis range of 0.1 to 1 AU.  
The combined binary fraction is large even without
accounting for systems in the relatively unexplored 
separation range of 1 to 100 AU.  The derived multiplicity 
fraction is $MF = 0.65 \pm 0.05$ and the companion frequency is
$CF = 1.11 \pm 0.13$.  We emphasize that these are lower limits
to the actual fractions because our observations miss
both very close and faint companions and because the fainter
companions are indistinguishable from background stars.
Nevertheless, our results are broadly consistent with
earlier surveys of massive stars that include both 
spectroscopic (close) and resolved (wide) binaries.
For example, the {\it HST}/FGS survey of O-type stars by \citet{ald15}
yielded $MF = 0.51 - 0.69$ and $CF = 0.70 - 1.67$ among cluster
and association stars, and the VLTI/PIONIER and
NACO/Sparse Aperture Masking survey of O-stars by \citet{san14}
led to $MF = 0.91 \pm 0.03$ and $CF = 2.2 \pm 0.3$.
This very high incidence of bound companions is consistent
with the idea that massive star formation directs the 
angular momentum of the natal cloud into the creation 
of binary orbital motion. 

The NIRI survey will help in the selection of targets for
future adaptive optics and integral field unit spectroscopy
observations to determine the physical properties of the companions.
The close companion stars detected in the NIRI survey are especially
interesting because their flux is blended into that of the
main target for most ground-based observations that lack
high angular resolution.  Thus, the NIRI results can help
correct the placement of these stars in the H-R diagram and
can inform the interpretation of spectroscopy of hierarchical
triples and other composite spectrum targets (see the case
of MT~429; \citealt{kim12a}).  Finally, the closest resolved
binaries hold the potential for orbital solutions and mass
determination of the most massive stars. For example, 
S 5 = Cyg~OB2 \#5 is an hierachical system consisting
of a central massive close binary, nearby tertiary, plus the 
two distant resolved companions \citep{rau19}. 
The brightest and presumably most massive star 
in Cyg~OB2 is MT~304 = Cyg~OB2 \#12,
and both the close companion found by {\it HST}/FGS and
the more distant companion found in the NIRI survey
were detected in speckle observations by \citet{mar16},
who claim that the close component has already displayed some
orbital motion.  The orbital period is probably $P\approx 100$~yr,
so continued high angular resolution observations hold
the promise of weighing the most massive star in Cyg~OB2
and one of the most massive stars in the Galaxy.


\acknowledgments

Based on observations obtained at the Gemini Observatory,
(GN-2005B-Q-64, GN-2008A-Q-85, GN-2008B-Q-95) which is operated by the
Association of Universities for Research in Astronomy, Inc., under a
cooperative agreement with the NSF on behalf of the Gemini
partnership: the National Science Foundation (United States), National
Research Council (Canada), CONICYT (Chile), Ministerio de Ciencia,
Tecnolog\'{i}a e Innovaci\'{o}n Productiva (Argentina), Minist\'{e}rio
da Ci\^{e}ncia, Tecnologia e Inova\c{c}\~{a}o (Brazil), and Korea
Astronomy and Space Science Institute (Republic of Korea). We thank
the staff of the Gemini North Observatory and especially Dr.\ Andrew
Stephens for their support of this program. The data was processed
using the Gemini IRAF package.  This paper contains observations
obtained at the Hale Telescope, Palomar Observatory. This research has
made use of the Washington Double Star Catalog maintained at the
U.S.\ Naval Observatory.  This publication also made use of data
products from the Two Micron All Sky Survey, which is a joint project
of the University of Massachusetts and the Infrared Processing and
Analysis Center/California Institute of Technology, funded by the
National Aeronautics and Space Administration and the National Science
Foundation.  This work was directly supported by the National Science
Foundation under grants AST-1009080 and AST-1411654.  Institutional
support has been provided from the GSU College of Arts and Sciences
and from the Research Program Enhancement fund of the Board of Regents
of the University System of Georgia, administered through the GSU
Office of the Vice President for Research and Economic Development. 
A portion of the research in this paper was carried out at the Jet
Propulsion Laboratory, California Institute of Technology, under a
contract with the National Aeronautics and Space Administration (NASA). Financial support was provided to SCN by the Science and Technology
Facilities Council for part of this work at the University of
Sheffield, with current support provided by the Florida Institute of
Technology. We are grateful for all this support.

{\it Facilities:} \facility{Gemini (NIRI)}
\software{FITSTARS \citep{ten00}, SEextractor \citep{ber96}} 



\appendix

\section{NIRI/ALTAIR Astrometry Calibration}

Each NIRI observation comes with World Coordinate System (WCS) information
in the FITS header that is retained through the image reduction and 
coaddition process.  These keywords list the pointing position and the 
right ascension and declination changes with pixel spacing along both axes. 
In principle, these can be used with the $(x,y)$ positions of stars in the 
merged image that were measured with SExtractor to derive the celestial 
coordinates $(\alpha, \delta)$.  However, there are several complications 
that need to be considered.  First, the pixel scale changed with the 
introduction of the field lens according to the Gemini Web 
site\footnote{http://www.gemini.edu/sciops/instruments/niri/imaging/pixel-scales-and-fov}
(see Table~\ref{t7-niri}), but this change was neglected in the WCS header keywords.  
Second, there is an apparent barrel distortion in the NIRI $f/32$ camera images
that causes stars at the periphery to appear closer to the center than 
they should based upon a strict linear plate scale (see notes at 
the Gemini Web site).  Finally, it is important to make an independent
check on the field rotation parameter in the FITS header.  

We decided to verify the pixel scales and rotational zero point through
a comparison of the relative $(x,y)$ positions with astrometry of the 
targets from the UK Infrared Telescope Infrared Deep Sky Survey 
(UKIDSS; \citealt{law07}).  The celestial coordinates in UKIDSS (J2000 equinox)
are based upon stellar positions in the 2MASS survey \citep{skr06} 
and hence are indirectly related to the International Reference Coordinate
System through the Tycho-2 system used by 2MASS \citep{lod07}.


Our goal was to obtain plate solutions for the field rotation and the
$x$- and $y$-axis pixel scales from our $(x,y)$ positions and the 
corresponding UKIDSS $(\alpha, \delta)$ coordinates for as many fields 
as possible.  The first step was to remove the barrel distortion effects. 
We assumed that the main target occupied the axial central position, 
and that the radial distance $r$ of any other star from the image center 
equals the uncorrected linear distance from the main target. 
However, this is an approximation, because the dither pattern placed 
the target in the center in only one of the nine dither locations, and the 
star is displaced by 0 or $\pm 50$ pixels in $x$ and $y$ for the other 
dither placements.  In fact, the distortion correction should actually be made
before image coaddition to avoid variations in radial distance between the 
target and image center in the individual frames, but the dither offsets are
small enough for our observations that the positional smearing that results 
from coaddition before barrel distortion correction only amounts to about 
one pixel at the edge of the FOV.  The true radial distance corrected for barrel distortion is 
$$r^\prime = r +k r^2$$ 
where $k= (1.32 \pm 0.02)\times 10^{-5}$ and $r$ is given in pixels (see Gemini Web site). 
Then the relative position from center $(\triangle x, \triangle y)$ may be 
transformed to a barrel distortion corrected position at  
$$\triangle x^\prime = \triangle x (r^\prime / r) = 
 \triangle x + k \triangle x (\triangle x^2 + \triangle y^2)^{1/2}$$
and 
$$\triangle y^\prime = \triangle y + k \triangle y (\triangle x^2 + \triangle y^2)^{1/2}.$$

Next we obtained UKIDSS $K$-band source data for the nominal position of 
the main target (from 2MASS) using a 15 arcsec search 
radius\footnote{http://surveys.roe.ac.uk:8080/wsa/region\_form.jsp}. 
The stellar positions were extracted from the UKIDSSDR7PLUS 
data release of the UKIDSS Galactic Plane Survey. 
We used the preliminary WCS header data to transform 
$(x,y)$ to $(\alpha, \delta)$ to then match our targets with the sources 
in UKIDSS (where possible) based upon similar coordinates and magnitudes.  
Finally, we used the positional and coordinate data to obtain a plate solution 
using the IDL procedure astromit.pro (written by R.\ Cornett and 
W.\ Landsman\footnote{http://www.astro.washington.edu/docs/idl/idllib/obsolete/sunuit/lib/old/astromit.pro}). 
The results for each field were collected in a file 
that listed the rotation angle and pixel scale in $x$ and $y$ for both the 
preliminary WCS data and the fit of the UKIDSS coordinates, plus the number of 
stars used in the fit.  

We found that there were 49 fields where four or more stars were matched by
sources in UKIDSS, and we used these to determine mean values of the 
pixel scales and rotational offsets that are summarized in Table \ref{t7-niri} 
according to the field lens position (out for the 2005 observations and
in for those from 2008).  The first two rows give the expected pixel 
scales from the Gemini Web site and the work of \citet{sto06}, and 
the next two rows show the average of the $x$ and $y$ pixel scales according to 
the preliminary WCS keywords and the fit of the UKIDSS astrometry, respectively. 
The uncertainties quoted are the standard deviations of the mean in each case.   
We see that the pixel scales are close to the expected values, and the 
ratio of the fitted to WCS pixel scales (given in the fifth row 
as the WCS scale factor) is slightly less than one.  Finally, there is 
a small but non-zero offset between the field rotational angle $\theta$ 
from the preliminary WCS keyword and the fits of the UKIDSS astrometry,
$\triangle\theta = \theta ({\rm UKIDSS}) - \theta ({\rm WCS})$. 
 
\begin{deluxetable}{lcc} 
\tablewidth{0pc} 
\tablenum{7} 
\tablecaption{Astrometric Scales for the NIRI/ALTAIR $f/32$ Camera\label{t7-niri}} 
\tablehead{ 
\colhead{Parameter}         & 
\colhead{Field lens out}    & 
\colhead{Field lens in} 
} 
\startdata 
Pixel scale [Gemini WWW] (mas pix$^{-1}$)  &  21.9              &  21.4              \\
Pixel scale [Stoesz 2006] (mas pix$^{-1}$) & $21.8 \pm 0.2$     & $21.4\pm  0.2$     \\
Pixel scale [WCS] (mas pix$^{-1}$)         & $21.859\pm 0.012$  & $21.860\pm 0.003$  \\
Pixel scale [fit] (mas pix$^{-1}$)         & $21.781\pm 0.025$  & $21.298\pm 0.008$  \\
WCS scale factor from fit                  & $0.9964\pm 0.0013$ & $0.9743\pm 0.0004$ \\
$\triangle\theta$ (deg)                    & $0.59\pm 0.12$     & $0.40\pm 0.03$     \\
Number of fields                           &  6                 &        43          \\
\enddata 
\end{deluxetable} 

\clearpage


We used these calibration results to determine the $(x,y)$ to $(\alpha, \delta)$
transformation using the IDL procedure 
xyad.pro\footnote{http://idlastro.gsfc.nasa.gov/ftp/pro/astrom/xyad.pro} 
(written by W.\ Landsman)
that we modified by performing the barrel distortion correction (see above), 
making a small rotation of the $(\triangle x^\prime , \triangle y^\prime)$
positions using $\triangle\theta$ for the lens in/out solutions in Table~\ref{t7-niri}, 
and then rescaling the WCS pixel scales using the WCS scale factors 
for the lens in/out solutions in Table~\ref{t7-niri}. 
The relative coordinates were then transformed to absolute $(\alpha, \delta)$
using the 2MASS coordinates for the main target (J2000 equinox and ignoring 
the effects of proper motion between the times of the 2MASS survey and our observations). 
We caution that in some cases the 2MASS coordinates may actually represent
the center of light position between the main target and a close companion, so 
that in such cases all the $(\alpha, \delta)$ estimates may have systematic offsets. 
The relative positions $(\rho, \theta)$ should be regarded as our fundamental astrometric measurements.
The target MT421 (Cyg~OB2-22) was also observed by \citet{mai10}
using the Advanced Camera for Surveys (ACS) High Resolution Camera (HRC) on 
{\it Hubble Space Telescope} (red F850LP filter), and we compared the 
separations and position angles of the companions observed with {\it HST} and our 
calibrated astrometry to verify our calibration process.  We found that 
the mean difference in fractional separation for five companions was 
$0.0011\pm 0.0023$ and the mean difference in position angle was $0.11 \pm 0.15$ deg. 
Thus, our calibration of the astrometry leads to pixel scales that agree 
at the $0.1\%$ level and to systematic rotational differences at the $0.1$ deg level.  The standard deviation between the rectilinear 
positions from the {\it HST} and NIRI astrometry is about 0.008 arcsec 
for these five companion stars, and this may represent the magnitude of
any high-order geometric distortions that may exist in the NIRI ALTAIR 
astrometry system.



\section{NIRI/ALTAIR Photometry Calibration}

\begin{figure} 
\begin{center} 
{\includegraphics[height=15cm]{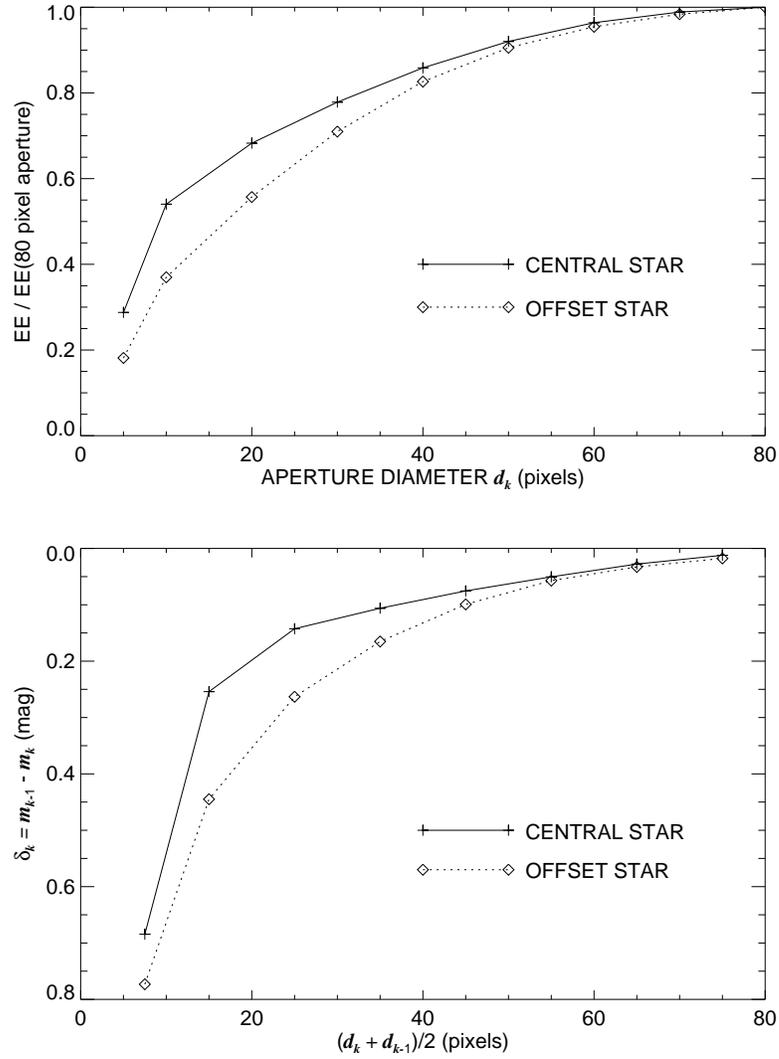}} 
\end{center} 
\caption{Above: A plot of encircled energy (EE) versus aperture diameter for MT~465 (solid line)
and a companion star near the edge (dotted line).  Both are normalized by the encircled 
energy for an aperture with a 80 pixel diameter.  Below: The difference in measured 
instrumental magnitude between sequential apertures of diameters $d_{k-1}$ and $d_{k}$ 
plotted versus the mean of these diameters. 
\label{f4-niri}} 
\end{figure}

The NIRI/ALTAIR images suffer from angular anisoplanatism that causes the 
point-spread function (PSF) to change from the center to the edge of the image. 
Stellar images near the periphery have relatively more flux in the halo surrounding 
the core than does a star image at the center.  We measured stellar fluxes using
aperture photometry with SExtractor for a series of apertures with diameters
ranging from 5 to 80 pixels, and these represent a radial integration of the 
stellar PSF.   Figure~\ref{f4-niri} shows an encircled energy (EE) plot of total 
flux measured versus aperture diameter for the case of a 2005 $K$-band observation
of MT~465.  The solid line represents the EE curve for the target at the 
center of image and the dotted line shows the EE curve for another star offset 
by 431 pixels from the main target.   We see that the PSF degradation of the 
offset star image results in a relative reduction in measured flux that is larger 
at smaller aperture size.  Consequently, if we adopted a fixed aperture diameter
of say 10 pixels for all our measurements, then we would systematically 
underestimate the flux of stars towards the edge of the field.  On the other
hand, if we used a larger diameter aperture (60 pixels or larger), then the 
differences in the EE flux with position would be insignificant.  Unfortunately, 
the large aperture option is only practical with the brightest and isolated stars 
because the stellar signal becomes overwhelmed by background noise for 
faint stars measured with large apertures (often causing the EE curve 
to decline with increasing aperture; \citealt{how89}).  Hence, we must apply 
an aperture correction scheme that accounts for the PSF degradation of   
our measurements.  

\input{epsf}
 
The amount of PSF degradation depends on the radial position of the star, 
the jitter introduced by the coaddition of the individual frames, 
whether or not the NIRI field lens was used, and the air mass
and seeing conditions at the time of the observation. 
\citet{cre05} argue that a first order correction can be 
made for PSF degradation by considering a family of PSFs characterized 
by the ratio of the offset angle to the isoplanatic angle (dependent on
air mass and seeing).  Because the isoplanatic angle is inversely proportional
to the astronomical seeing, this suggests that we may parameterize the 
changes in the EE curves using a parameter $\alpha \equiv r \theta_s$, 
where $r$ is the offset position from the target at the center of the image
(measured in pixels) and $\theta_s$ is the FWHM of the astronomical seeing 
(recorded in the NIRI/ALTAIR header files as keyword {\tt AOSEEING}).  
We first tested this idea by calculating EE curves for synthetic PSFs for 
NIRI/ALTAIR created with the PAOLA software package \citep{jol10}, and 
we found that the ratio $EE_k / EE_k[{\rm REF}]$ (where $EE_k$ is the normalized 
enclosed energy for aperture $k$ and $EE_k[{\rm REF}]$ is the same for the 
main target at center) did indeed decline in an approximately
linear fashion with both increases in radial offset and seeing.  However, 
the observed PSFs have sufficiently different core structure from the model PSFs 
(presumably due to jitter that is not included in the models) that we decided
to calibrate the change in the EE curves directly from our observations. 

We implemented the aperture correction using the differential 
magnitude approach outlined by \citet{ste90} in which the instrumental 
magnitude difference between two apertures is 
$$\delta_k = -2.5 \log (F_k / F_{k-1})$$
where $F_k$ is the flux estimated by SExtractor for an aperture of 
diameter $d_k$.  Uncertainties in $\delta_k$ were set by the flux uncertainties
according to the $S/N$ from equation (1) of \citet{how89}.  This differential 
version of the EE curve is shown in the lower part of Figure~4.  The advantange
of using the differential form $\delta_k$ is that this magnitude difference may be 
estimated for the smaller apertures even for those faint stars where the EE curve is 
unreliable at larger apertures because of background noise \citep{ste90}.  

We then gathered $\delta_k$ measurements for all the aperture pairs
for image samples selected by date (to account for the use or not of
the field lens) and by filter band.  In each case we formed the
difference between $\delta_k$ for a given star and that for the
central reference target, and we collected the offset parameter
$\alpha = r \theta_s$.  The uncertainties in $\alpha = r \theta_s$ are
estimated as $\pm 15\%$, which reflects the typical scatter in seeing
estimates among the subexposures.  An uncertainty weighted fit was
made of the function
$$\delta_k - \delta_k({\rm REF}) = a_k \alpha$$ for the first-order
model of PSF degradation with parameter $\alpha$.  The derived
constants $a_k$ and their uncertainties are collected in Table
\ref{t8-niri} for each year and filter sample.  The second row in the
header indicates the associated aperture pair (by diameter in pixels)
for each column.   The PSF degradation trends are largest in the smaller
aperture pairs, shorter wavelength filters, and data from 2005 when
the AO field lens was not used.  We also list in Table \ref{t8-niri} similar
coefficients for the PSF degradation observed in the PHARO images
(made in 2009), but these should not be directly compared with the
NIRI/ALTAIR results because the pixel scale is different and no seeing
estimate was reported at the time, but based on AO performance, we
estimate the data were taken in approximately $0.8 \arcsec$ seeing
\citep{dek07}. 

\begin{deluxetable}{lcccccccc} 
\tablewidth{0pc} 
\tabletypesize{\scriptsize}
\tablenum{8} 
\tablecaption{PSF Degradation Correction Coefficients and Approximation Uncertainties\label{t8-niri}} 
\tablehead{ 
\colhead{Image   }          & 
\colhead{$a_1 \times 10^6$} & 
\colhead{$a_2 \times 10^6$} & 
\colhead{$a_3 \times 10^6$} & 
\colhead{$a_4 \times 10^6$} & 
\colhead{$a_5 \times 10^6$} & 
\colhead{$a_6 \times 10^6$} & 
\colhead{$a_7 \times 10^6$} & 
\colhead{$a_8 \times 10^6$} \\
\colhead{Set}               & 
\colhead{ (5--10)}          & 
\colhead{(10--20)}          & 
\colhead{(20--30)}          & 
\colhead{(30--40)}          & 
\colhead{(40--50)}          & 
\colhead{(50--60)}          & 
\colhead{(60--70)}          & 
\colhead{(70--80)}            
} 
\startdata 
2005 $J$  \dotfill &$1130\pm 46$&$1095\pm 28$&$169\pm 20$&$ 65\pm 18$&$ 30\pm 17$&$-11\pm 16$&$-25\pm 16$&$-32\pm 16$\\
2005 $H$  \dotfill &$ 677\pm 47$&$ 960\pm 33$&$473\pm 27$&$104\pm 25$&$ 34\pm 24$&$ 15\pm 24$&$ -2\pm 26$&$  0\pm 28$\\
2005 $K$  \dotfill &$ 287\pm 50$&$ 770\pm 35$&$465\pm 29$&$223\pm 27$&$104\pm 32$&$ 40\pm 35$&$ -5\pm 41$&$  7\pm 27$\\
2008 $J$  \dotfill &$ 976\pm 23$&$ 896\pm 15$&$251\pm 12$&$ 72\pm 10$&$ 22\pm  9$&$-53\pm  9$&$-67\pm 10$&$-42\pm 10$\\
2008 $K$  \dotfill &$   4\pm 19$&$ 510\pm 14$&$225\pm 14$&$ 69\pm 14$&$ 39\pm 14$&$ 17\pm 16$&$-20\pm 16$&$-33\pm 15$\\
2009 $J$  \dotfill &$ 166\pm 25$&$ 326\pm 18$&$110\pm 16$&$ 44\pm 16$&$ 81\pm 14$&$ 23\pm 12$&$ 17\pm 16$&$ 32\pm 17$\\
2009 $H$  \dotfill &$ 289\pm 17$&$ 297\pm 11$&$112\pm 10$&$ 28\pm 10$&$ 10\pm  9$&$-23\pm  9$&$ -5\pm  9$&$ 16\pm  9$\\
2009 $K_S$\dotfill &$ 176\pm 15$&$  56\pm 10$&$ 84\pm  9$&$ 25\pm  9$&$  4\pm  9$&$  0\pm  9$&$ -1\pm  9$&$-20\pm  9$\\
\tableline
$\sigma(\delta_k[{\rm MODEL}])$ (mag)
          & 0.162      & 0.049      & 0.022     & 0.010     & 0.009     & 0.010     & 0.014     & 0.017     \\
\enddata 
\tablecomments{Any negative values are assigned zero in practice.}
\end{deluxetable} 

\clearpage

We used the observed PSF degradation trends to estimate a model differential 
magnitude curve $\delta_k$ for each target's position according to  
$$\delta_k = \delta_k[{\rm REF}] + a_k \alpha$$
where $\delta_k[{\rm REF}]$ is the magnitude difference between apertures 
$k-1$ and $k$ for the central reference star, $a_k$ is the coefficient for 
a given date and filter (given in Table \ref{t8-niri}), and $\alpha = r \theta_s$ is 
the radial distance -- seeing product.  It is important to check how well 
this approximate treatment works in practice, so we compared the 
predicted curve $\delta_k[{\rm MODEL}]$ with those observed for a subsample 
of 16 very bright and radially offset stars where the uncertainties due to 
photon and background noise are insignificant.  The standard deviations 
of the observed minus model $\delta_k$ curves, $\sigma(\delta_k[{\rm MODEL}])$, 
are given with each entry in Table \ref{t8-niri}, and these represent how well 
we might expect the model to perform in our application.  In general 
these standard deviations are small but they are worse for the smallest 
aperture pairs where structure variations in the PSF are most pronounced.
The full uncertainty in our $\delta_k$ estimate is given by 
$$\sigma^2(\delta_k) = \sigma^2(\delta_k[{\rm MODEL}]) + \sigma^2(\delta_k[{\rm REF}])
   +  \sigma^2(a_k \alpha)$$
where the final term accounting for the off-axis correction is 
$$\sigma^2(a_k \alpha) = 
   (a_k \alpha)^2\left(\left({{\sigma(a_k)}\over{a_k}}\right)^2+\left({{\sigma(\alpha)}\over{\alpha}}\right)^2\right) 
   \approx \alpha^2\left(\sigma(a_k)^2+a_k^2\left({{\sigma(\theta_s)}\over{\theta_s}}\right)^2\right).$$
The approximation used in the last step assumes that all the uncertainty in the $\alpha = r \theta_s$
product stems from the seeing uncertainty $\sigma(\theta_s)$.  The uncertainties in the 
coefficients $\sigma(a_k)$ are given with each entry of Table \ref{t8-niri}, 
and we adopt $\sigma(\theta_s)/\theta_s = 0.15$.

Now with the off-axis aperture curves $\delta_k$ in hand, we may estimate 
the magnitude difference between target and central reference star using an 
aperture correction as given by 
$$\triangle m_k = -2.5 \log (F_k / F_{80}[{\rm REF}]) - \sum_k^8 \delta_k$$
where we refer all the fluxes to that in the largest, 80 pixel diameter aperture 
of the reference star.  The uncertainty associated with this magnitude difference is 
$$\sigma^2(\triangle m_k) = \sigma^2(F_k) + \sum_k^8 \sigma^2(\delta_k)$$ 
where $\sigma(F_k)$ is the uncertainty in the flux measurement expressed as a
magnitude and $\sigma(\delta_k)$ are the uncertainties in the adopted $\delta_k$
curve as given above.  Thus, we arrive at nine estimates of the
magnitude difference and associated uncertainty from the measurements
made in nine apertures.  We select the estimate with the smallest
uncertainty for our purposes in this paper, so that we can adopt the
best compromise between large apertures for the bright stars (where
the flux uncertainties are small compared to the aperture correction
uncertainties) and smaller apertures for the fainter stars (where the
flux uncertainties become huge in the large apertures). Note that
stars at the periphery of the fields (i.e. stars that were not in all
frames due to dithering) will have larger uncertainties than reported
in Table~\ref{t4-niri}.

We checked our scheme by comparing our derived differential $K$-band magnitudes with those
from the UKIDSS catalog for the populous field surrounding star MT~421.  The individual 
stars were matched between the NIRI and UKIDSS sources according to our astrometry solution.
Unfortunately, MT~421 itself is saturated in the UKIDSS data, so it is not possible to 
form magnitude differences from the UKIDSS data alone.  Instead, we found the best 
fit magnitude offset needed to match the NIRI magnitude differences, and 
the implied $K$ magnitude of MT~421 is $K=7.77$ which is similar to the estimate from 
2MASS, $K=7.76$.  We find that our corrected magnitudes and those from UKIDSS 
are in satisfying agreement with no evidence of systematic differences with magnitude. 
Furthermore, the scatter about the expected one-to-one relation
is comparable to our uncertainty estimates, which suggests that our analytical 
representation of the uncertainties is reliable.  



\bibliographystyle{apj}
\bibliography{apj-jour,paper}


\end{document}